\documentclass[prd,aps,10pt,twocolumn,floatfix,nofootinbib,superscriptaddress]{revtex4-1}
\pdfoutput=1
\usepackage{lipsum}
\usepackage[utf8x]{inputenc}
\usepackage{textcomp}
\usepackage{amsmath}
\usepackage{amsfonts}
\usepackage{amssymb}
\usepackage{mathrsfs}
\usepackage{graphicx}
\usepackage{hyperref}
\usepackage[usenames, dvipsnames]{color}
\usepackage[retainorgcmds]{IEEEtrantools}
\usepackage{gensymb}
\usepackage{verbatim}
\usepackage[caption=false]{subfig}

\newcommand{\onion}{{\sc Onion} }
\newcommand{\skylight}{{\sc Skylight}}

\begin{document}

\title{Relativistic force-free models of the thermal X-ray emission in millisecond pulsars observed by NICER}

\author{Federico Carrasco}
\email{fcarrasc@famaf.unc.edu.ar}
\affiliation{Instituto de F\'isica Enrique Gaviola, CONICET, Ciudad Universitaria, 5000 C\'ordoba, Argentina}
\affiliation{Facultad de Matem\'atica, Astronom\'ia, F\'isica y Computaci\'on, Universidad Nacional de C\'ordoba, Argentina}

\author{Joaquin Pelle}
\email{jpelle@mi.unc.edu.ar}
\affiliation{Instituto de F\'isica Enrique Gaviola, CONICET, Ciudad Universitaria, 5000 C\'ordoba, Argentina}
\affiliation{Facultad de Matem\'atica, Astronom\'ia, F\'isica y Computaci\'on, Universidad Nacional de C\'ordoba, Argentina}

\author{Oscar Reula}
\email{reula@famaf.unc.edu.ar}
\affiliation{Instituto de F\'isica Enrique Gaviola, CONICET, Ciudad Universitaria, 5000 C\'ordoba, Argentina}
\affiliation{Facultad de Matem\'atica, Astronom\'ia, F\'isica y Computaci\'on, Universidad Nacional de C\'ordoba, Argentina}

\author{Daniele Vigan\`o}
\email{daniele.vigano@csic.es}
\affiliation{ Institute of Space Sciences (ICE, CSIC), 08193 Barcelona, Spain}
\affiliation{Institut d'Estudis Espacials de Catalunya (IEEC), 08034 Barcelona, Spain}
\affiliation{Departament  de  F\'{\i}sica $\&$ IAC3, Universitat  de  les  Illes  Balears,  Palma  de  Mallorca,  Baleares  E-07122,  Spain}

\author{Carlos Palenzuela}
\email{carlos.palenzuela@uib.es }
\affiliation{Departament  de  F\'{\i}sica $\&$ IAC3, Universitat  de  les  Illes  Balears,  Palma  de  Mallorca,  Baleares  E-07122,  Spain}
\affiliation{Institut d'Estudis Espacials de Catalunya (IEEC), 08034 Barcelona, Spain}

\date{\today}

\begin{abstract}

Several important properties of rotation-powered millisecond pulsars (MSPs), such as their mass-radius ratio, equation of state and magnetic field topology, can be inferred from precise observations and modelling of their X-ray light curves. In the present study, we model the thermal X-ray signals originated in MSPs, all the way from  numerically solving the surrounding magnetospheres up to the ray tracing of the emitted photons and the final computation of their light curves and spectra. Our modelled X-ray signals are then compared against a set of very accurate NICER observations of four target pulsars: PSR J0437–4715, PSR J1231−1411, PSR J2124−3358 and PSR J0030+0451. We find very good simultaneous fits for the light curve and spectral distribution in all these pulsars.

The magnetosphere is solved by performing general relativistic force-free simulations of a rotating neutron star (NS) endowed with a simple centered dipolar magnetic field, for many different stellar compactness and pulsar misalignments.
From these solutions, we derive an emissivity map over the surface of the star, which is based on the electric currents in the magnetosphere. In particular, the emission regions (ERs) are determined in this model by spacelike four-currents that reach the NS.
We show that this assumption, together with the inclusion of the gravitational curvature on the force-free simulations, lead to non-standard ERs facing the closed-zone of the pulsar, in addition to other ERs within the polar caps. The combined X-ray signals from these two kinds of ERs (both antipodal) allow to approximate the non-trivial interpulses found in all the target MSPs light curves.   

\end{abstract}


\maketitle

\section{Introduction}

Pulsars were first discovered in the 1960s, and knowledge about their nature and physical properties has been growing relentlessly ever since. In particular, the last two decades have witnessed remarkable progress both in theoretical understanding and observational capabilities. However, despite great advances, some key aspects are yet to be fully understood. The neutron star (NS) interior compositions and equations of state, the magnetic field configuration, and the structure of the surrounding plasma remain fundamental open questions in pulsar astrophysics.

Rotation-powered millisecond pulsars (MSPs) are a peculiar class of pulsars with highly stable spin rates. Their periods are around $1-30\,$ms, and their spin-down rates are of the order of $\dot{P} \sim 10^{-20}$. According to the usual magnetic braking model, they are old objects with characteristic ages $\tau_c = P/2\dot{P} \sim 0.1 - 10\,$ Gyr, and they have relatively low surface magnetic fields $B \sim 10^{8-9}\,$ G. Outside of the radio band, the greatest amount of information about these objects can be found in the soft X-ray range (∼0.1–10 keV). The emission in this range is mainly of thermal origin, although non-thermal components (usually modeled by power laws) often exist as well. The thermal component is produced in small hot regions on the surface of the star, commonly called ``hot spots''. These regions are arguably heated to temperatures up to $T \sim 10^6\,$ K by the constant bombardment of particles that accelerate in magnetospheric gaps up to very high energies \cite{ruderman1975}.

A good method for indirectly measuring several important properties of MSPs is the precise observation and modelling of their X-ray light curves \cite{pechenick1983hot, strohmayer1992light, page1994, miller1998effects, braje2000light, harding2001, harding2002, beloborodov2002gravitational, cadeau2007light, morsink2007oblate, zavlin2007studying, bogdanov2007, bogdanov2008, ozel2012, psaltis2014pulse, psaltis2014prospects, miller2015determining, nattila2018radiation, baubock2019, lockhart2019x, bilous2019, riley2019, bogdanov2019constrainingI, bogdanov2019constrainingII, miller2019, salmi2020, chen2020, kalapotharakos2021}. MSPs are especially suited for this approach because of their highly stable spin rates and pulse shapes.
These techniques have been applied, with increasing level of refinement, for four decades already. However, despite the sophistication achieved, pulsar X-ray light curve fitting continues to be a highly nontrivial task, and the problem is in general quite degenerate unless very good observational data is present.

One important application of pulsar pulse profile modelling to the inference of physical properties is in constraining their mass-radius relations and so the fundamental equation of state. The latter closes the system of stellar structure equations: it is therefore a necessary ingredient to have the solution of the star and the spacetime around it, once the central energy density and the rotation rate have been fixed. In this manner, a mapping is obtained which translates equations of state into mass-radius relations. Given that thermal X rays are emitted over the surface of the star, the stellar compactness has a strong impact on the luminosity profiles through the gravitational field in which the rays propagate. It is for that reason that pulse profile modelling can be used to pose constraints on compactness, and, via the inverse mapping, on the equations of state. Many recent theoretical and observational efforts have been directed towards that end. Likewise, the technique can be used to determine the geometry of the magnetic field close to the surface, because the location and shape of the hot spots depend on it, which in turn determines the observed luminosity profiles.

As mentioned before, the pulse profiles are subject to various physical effects. The impact of the gravitational field occurs through the deflection of light rays, and the gravitational redshift of the photon frequencies. Additionally, the spin of the star also affects the frequency of emitted light through local Doppler boosting according to the tangential speed of the emitting portion of the surface. These effects need to be taken into account by relativistic ray tracing and radiative transport calculations. For rapidly rotating neutron stars, the gravitational field has to be numerically computed as part of the solution of the stellar structure equations. However, for rotation frequencies below 400 Hz, the Kerr spacetime is a sufficient approximation of the exterior gravitational field of a NS \cite{miller1998effects, braje2000light}.

The Neutron Star Interior Composition Explorer (NICER; \cite{gendreau2016neutron}) began operating in June 2017, and has since measured the soft X-ray light curves of a few nearby rotation-powered MSPs with unprecedented sensitivity (e.g., \cite{bogdanov2019constrainingI, guillot2019nicer}). One of its main purposes is, precisely, to obtain high quality data for inferring the mass and radii of the observed neutron stars, and the geometry of their surface magnetic fields, via pulse profile modelling.

 A Bayesian modelling of the X-ray light curve has been performed for PSR J0030+0451 \cite{miller2019, riley2019, bilous2019}, one of the pulsars observed by NICER. They constructed multi-parametric families of models with two circular or annular hot spots of uniform temperature, allowing for non-antipodal configurations. Then, they did a statistical sampling over the parameters of the families, searching for the best fits to the light curve. They found that the sampling favoured the non-antipodal configurations, suggesting that the surface magnetic field of PSR J0030+0451 should have a strong non-dipolar component. Posterior works \cite{chen2020, kalapotharakos2021} searched for vacuum magnetic field configurations such that the foot-prints on the stellar surface of those  
field lines crossing the light cylinder would closely resemble the hot spots suggested by the observations; this included offset dipole-plus-quadrupole components. Then, they seeded these vacuum fields to force-free simulations to obtain the polar caps, defined as the regions on the surface where the magnetic field lines are open. 
Using this information on the global magnetospheric structure, and assuming a uniform temperature over these polar caps, they were able to construct multi-wavelength emission models, including radio and $\gamma$-rays, that were consistent --to some extent-- with observations. 

In this paper, we pursue a slightly different route. We start from general relativistic force-free (GRFF) simulations of a rotating NS, endowed with a standard centered dipolar magnetic field configuration (as done in \cite{Pulsar}); and then, we link the resulting electric currents on the surrounding plasma with the surface thermal emission from the star, essentially following the model proposed in \cite{lockhart2019x}. Once we have the emissivity map from the NS surface, we perform the relativistic ray tracing with the recently developed code \skylight~\cite{Skylight}, to compute the corresponding light curves and spectra.
We observe that the additional input from the GRFF simulations in modelling the emissions results in more complex ERs than the traditional circular polar caps associated to the centered dipolar magnetic field. Here, we explore the possibility that such new structures for the emitting regions could help in accounting for the observational X-ray data of MSPs, without the need 
of invoking strong non-dipolar magnetic field components and/or offsets from the NS center. We concentrate our exploration on a set of four target pulsars, PSR J0437–4715, PSR J1231−1411, PSR J2124−3358 and PSR J0030+0451, for which there are good quality X-ray data publicly available (see \cite{bogdanov2019constrainingI} and references therein). We find that our approach allow to reproduce the observations with surprising accuracy.

The article is organized as follows. In Section \ref{sec:setup} we describe the methods employed, from the modelling of the pulsar magnetosphere up to the fitting of our modelled light curves and spectra to the observational data. In Section \ref{sec:results},
we first introduce the emission regions derived from the GRFF simulations, and then present and analyze our best-fits to the X-ray data of the four target MSPs. Additional information regarding the fitting procedure and relevant posterior probabilities distributions can be found in Appendix \ref{sec:appendix}. Finally, we summarize our work and conclude in Section \ref{sec:conclusions}.

\section{Methods}\label{sec:setup}

We shall tackle the problem by four separate instances. First, we numerically solve the magnetosphere surrounding a rotating pulsar endowed with a simple dipolar magnetic field. We use the force-free approximation, which provides with a global solution for the electromagnetic field. Then, we model a temperature map at the stellar surface,
by linking the magnetospheric currents to the energy deposited at the surface (or in the cm-thick atmosphere) by the bombardment of relativistic particles.
We then perform the ray tracing to simulate the detection of emitted photon by different distant observers and compute their corresponding light curves and spectra. Finally, we compare our models with the observations by NICER and XMM-Newton, searching for those configurations that better reproduce the X-rays light curves and spectra.

\subsection{Pulsar magnetosphere}

We employ the numerical code \onion to solve for the exterior force-free (FF) magnetosphere surrounding a neutron star. This code has been used to model several astrophysical scenarios involving black holes and neutron star magnetospheres (see \cite{FFE2, Pulsar, Boost, carrasco2019, carrasco2020, carrasco2021}) and, in particular, it has already been applied to study misaligned pulsars in general relativity \cite{Pulsar}. Hence, we briefly describe here the most relevant aspects of our setup for the purposes of the present work, while referring the reader to \cite{FFE2, Pulsar} for technical details about the numerical implementation. 

We approximate the exterior of a rotating neutron star by the Kerr metric, parametrized by the mass $M$ and spin $a$. 
Assuming a spherically symmetric neutron star of radius $R$,
we fix the dimensionless moment of inertia $\mathcal{I} := I/MR^2$ to the value $2/5$ (as in, e.g., \cite{petri2015general, gralla2016pulsar}). 
This allow to relate the star angular velocity $\Omega$ with the spin parameter in units of the stellar radius, namely:
\begin{equation}\label{eq:spin}
 \frac{a}{R}=\frac{2}{5} \, \Omega \, R
\end{equation}
Note that realistic stars are expected to have a slightly lower value $\mathcal{I}$, since the mass is concentrated towards the center. However, the value $2/5$ remains a good approximation within the range of typical NS compactness \cite{bejger02}.

In the present work, we consider a standard (centered) magnetic dipole field configuration, as we would like to test, in our model, the need for significant higher-multipoles components and/or off-centered fields to account for (some of) the observed light curves by NICER.

Within this setup, there are three dimensionless parameters which completely characterize a given GRFF simulation, namely: (i) the misalignment angle $\chi$ between the magnetic moment and the rotational axis of the pulsar; (ii) the surface rotation velocity $v_s := \frac{R}{R_{LC}} $, being $R_{LC}$ the light-cylinder radius; and (iii) the stellar compactness,
\begin{equation}\label{eq:compactness}
    \mathcal{C} := \frac{M}{R} \left(\frac{G}{c^2} \right)
\end{equation}
where $G$ is Newton's constant and $c$ the speed of light.

Notice that, once these dimensionless quantities are fixed, choosing the physical value of one of the three parameters (i.e., $R$, $M$ or $\Omega$) completely determine the other two.
On the other hand, we note that the magnetic field strength can be freely re-scaled\footnote{This is a well known property of the FF equations (e.g., \cite{komissarov2002}).} and, thus, can be given \textit{a-posteriori}. 
Meaning, in practice, that we can use the same FF run to represent pulsars of different magnetic field strengths, whose values are inferred from the accurate measurements of the timing properties (spin period $P$ and its time derivative $\dot{P}$), via the popular --and rough-- vacuum spin-down formula estimate, $B_{\rm pole} [G]=6.4 \times 10^{19} (P[s]  \dot{P})^{1/2}$ (where $B_{\rm pole}$ is the surface magnetic field strength at the pole). 

For the present study, we perform an initial set of 25 GRFF simulations at fixed rate $v_s = 0.05$ (which can represent, for instance, a pulsar with a $4.8\,$ms period and NS radius of $11.5\,$km), exploring various misalignments $\chi= \{ 15\degree, 30\degree, 45\degree, 60\degree, 75\degree \}$ and stellar compactness $\mathcal{C} = \{ 0.15, 0.18, 0.2, 0.22, 0.25 \}$ (which correspond to masses $M \sim \{ 1.2, 1.4, 1.6, 1.7, 1.9 \} \, M_{\odot}$ in the above example). 

We solve the system in a region between the NS surface at radius $R$, and an exterior spherical surface located at $64 R \gtrsim 3 R_{\rm LC}$.
The domain is represented by a total of $6 \times 6$ sub-domains, with $6$ patches to cover the angular directions, times $6$ spherical shells expanding in radius. These spherical shells do not overlap each other, and have more resolution in the inner regions: from layer to layer, the radial resolution is decreased by a factor $2$.
We adopt a resolution with grid numbers $N_{\theta} \times N_{\phi} \times N_{r}$ with $N_{\phi} = 2 N_{\theta} = 320$, while $N_{r} = 360$ to span the whole computational domain. 

In all cases, we evolve the system until it relaxes and reaches a stationary state (i.e., after roughly two stellar rotations) before extracting the FF field configuration. In particular, we look at the relevant information contained in the FF charge and current densities\footnote{As standard in FF electrodynamics, one can always recover partial information about the plasma from its four-current $J^a$, by means of the co-variant Maxwell equations, i.e.: $J^a = \nabla_b F^{ab}$.}, $\rho$ and $\vec{j}$, as seen by a co-moving observer rotating with the NS.

\subsection{Emission model} \label{sec:emission_model}

It is generally accepted that the thermal surface emission in millisecond pulsars arises from bombardment of relativistic particles hitting at the stellar surface. Such energetic particles are assumed to originate within the magnetosphere at the so-called \textit{gaps} where $E\cdot B \neq 0$ (i.e., regions where the force-free condition breaks-down). We shall follow here the approach presented in \cite{lockhart2019x},
in which pair-creation and accelerations are assumed to take place along a subset of the magnetospheric currents where the four-current $J^a$ is spacelike. The argument supporting this choice is that spacelike currents develop high voltage drops with counter-streaming flow of electrons and positrons, thus favoring intense bursts of pair-production \cite{timokhin2013, harding2022}.
 
Once the emitting region at the stellar surface is established by the above condition (i.e., $J_a J^a > 0$), the kinetic energy deposition is estimated by assuming that a fixed fraction $ \kappa $ of the spatial current $|\vec{j}|$ (defined in the local co-moving frame of the NS) is carried by relativistic particles with averaged Lorentz factor $\bar{\gamma}$, traveling towards the stellar surface (see \cite{baubock2019} for a detailed discussion). The rate at which this energy is deposited is balanced to the power radiated as a black body, yielding the temperature:
\begin{equation}\label{eq:Teff}
 T = \left( \frac{c^2 m_e \kappa (\bar{\gamma}-1)}{e \sigma} \right)^{1/4} |\vec{j}|^{1/4}   
\end{equation}
where $e$, $m_e$ and $\sigma$ represent the electron charge, electron mass and the Stefan-Boltzmann constant, respectively. 
The factor $\kappa$ relates to the pair multiplicity $\mathcal{M}$ through $\kappa = \zeta \mathcal{M}$, being $\zeta$ the ratio of in-going to total pairs \cite{salmi2020}. 
We note that since we are assuming relativistic particles, $\kappa (\bar{\gamma} -1) \approx \kappa \bar{\gamma} $, and it acts as a single effective parameter controlling the local temperature in the model.
Both $\kappa$ and $\bar{\gamma}$ (and, therefore, their product $\kappa \bar{\gamma}$), are largely unconstrained by the current theoretical understanding.

In order to convert this effective temperature to a model for the emerging specific intensity of the radiation field, one needs a detailed calculation of the thermal structure of the NS atmosphere, as it was done for instance in \cite{baubock2019, salmi2020}. Instead, we use a simplified approximation of blackbody emission at the effective temperature $T_{eff}$ (e.g., \cite{lockhart2019x}),
\begin{equation}
 I_{\nu} = \frac{\nu^3 \, \mathcal{B}(\Theta)}{\exp [\nu/T_{eff}] -1 } \label{eq:bb} 
\end{equation}
where $\nu$ is the frequency of the electromagnetic radiation and we have considered a non-isotropic beaming distribution given by,
\begin{equation}
 \mathcal{B}(\Theta) = \cos^b (\Theta)
\end{equation}
where $b$ is the anisotropy index and $\Theta$ the angle between the photon propagation direction and the surface normal. In previous works (e.g., \cite{kalapotharakos2021}), exponents in the range $b \sim 0.5-1.0$ have been used as a decent approximation to the NS atmosphere. This assumption is known to be needed to explain high pulsed fractions observed in the thermal X-ray light curves of many pulsars. 

Note that the simple analytical and energy-independent beaming correction is highly simplified, compared to the (energy-dependent) anisotropy obtained by detailed radiative models \cite{ozel13} (their Fig. 5, in particular). Moreover, besides introducing anisotropy, the atmosphere modifies the spectrum, so that, when it is modeled by a blackbody with an effective temperature $T_{eff}$ like in \eqref{eq:bb}, the latter looks higher than the physical temperature $T$. The ratio $T_{eff}/T$, called color-correction factor, is of the order $\sim 2$ for light-elements atmosphere \cite{sulemainov11}. Such factor, needed to convert the value of $T$, eq.~\eqref{eq:Teff}, to $T_{eff}$, is not explicitly indicated here, but it is implicitly reabsorbed in the free effective parameter $\kappa \bar\gamma$.

We will discuss the limitations of the simple emission model adopted here in Sections \ref{sec:normalization} and \ref{sec:conclusions}.

\subsection{Ray tracing and radiative transfer}

To transport the emission from the stellar surface to the observers, and thereby calculate light curves and spectra, we use the general-relativistic ray tracing and radiative transfer code \skylight~\cite{Skylight}, which supports arbitrary spacetime geometries. We adopt the full Kerr spacetime, with its spin and mass linked to the NS properties through equations \eqref{eq:spin}-\eqref{eq:compactness}. For this work, we use the observer-to-emitter scheme of the code, where we set a virtual image plane at the observer, and we take a bundle of vectors perpendicular to the image plane as initial tangent vectors to the geodesics. Once the initial data is set, the geodesic equations are integrated backwards in time, until the trajectories intersect the stellar surface.  

Then, for each observer we obtain a mapping from the image plane to the stellar surface, including the angles that the rays make with the surface normal. Since the spacetime is stationary, these mappings contain all the necessary information to compute the observed time-dependent fluxes. We compute these mappings for a single observer time. Then, for different times, we rotate the stellar surface to the corresponding phase and identify the rays intersecting the hot spots. For each ray, we consider the gravitational redshift and the Doppler boosting to the co-rotating frame of the star to calculate the specific intensity at the camera from the local emission model. Finally, we integrate over the image plane to obtain light curves and spectra.

In all the ray tracing simulations, the image planes are located at $750 r_g$, where $r_g = GM/c^2$, and we take grids of $600 \times 600$ rays over the image planes. These values are appropriate for our purposes, since locating the image plane further away, or increasing the resolution, do not significantly affect the results. For each pulsar, the simulations take the stellar compactness, $\mathcal{C}$, the spin parameter, $a$, and the viewing angle, $\xi$, as inputs: $\mathcal{C}$ and $a$ determine the spacetime geometry (we take units such that the stellar radius is fixed), and $\xi$ determines the position and orientation of the image plane. The rest of the parameters, as those listed in Table~\ref{table1}, only enter the calculation at the end, when retrieving the physical units of the calculated fluxes.

\subsection{Searching for the best-fits} \label{sec:search}

We want to find the set of parameters in our model that produces the best-fit to the observational data for the four target millisecond pulsars in \cite{bogdanov2019constrainingI}.
Ideally, one should directly compare the full phase-resolved spectral information against the observational data, as done in \cite{riley2019} for PSR J0030+0451. Our main aim here is to reproduce the light curves, for which NICER provides high-accuracy data. At the same time, we shall vary the parameters of our idealized emission model to reproduce the spectral shape as well.

We proceed as follows: we first estimate an appropriate value for the parameter $\kappa \bar{\gamma}$, which controls the effective temperature of the emitting regions in our model, by looking at the phase-averaged spectra. Then, we vary the remaining parameters --which only affect the spectrum very weakly-- in order to search for the light curve best-fits.
Once an approximated value for $\kappa \bar{\gamma}$ is found, it is fixed. Next, for each compactness considered by the GRFF simulations, we sample on observer orientation within the range $\xi \in (0, \pi/2)$ at intervals $\Delta \xi = 5\degree$, performing the ray tracing with \skylight. Thus, we get a discrete sample in $(\mathcal{C}, \chi, \xi)$, for which we are ready to compute their light curves, provided the value of the last parameter is given, i.e.: the anisotropy index $b$.

In order to find the best fits, we compare the curves normalized by their maximum flux values. That is, we compare $F/F_{max}$ and $F^{N}/F^{N}_{max}$, being $F$ and $F^N$ the modelled and measured (by NICER) fluxes, respectively. 
Notice that for the comparison one needs to account for the possible phase-shift $\varphi_o$ among the two curves, as well as a linear factor $\lambda_o$ arising from the fact that the peak values do not need to coincide (i.e., $F_{max}  = \lambda_o F^{N}_{max}$). Thus, what we really compare is $f(\varphi; \lambda_o, \varphi_o, b) := (\lambda_o F/F_{max}) (\varphi+\varphi_o; b)$ against $ f^N (\varphi) := (F^{N}/F^{N}_{max}) (\varphi)$ for the parameters $\{ \lambda_o, \varphi_o, b \}$.

The performance in the above comparison can be captured by the classical \textit{reduced chi-square} factor,
\begin{equation}\label{eq:dist}
 \chi_{r}^2 := \chi^2 / \varpi = \frac{1}{\varpi} \sum_{i}^{n}  \frac{\left( f(\varphi_i; \lambda_o, \varphi_o, b) - f^N (\varphi_i)\right)^2}{\varepsilon_{iN}^2} 
\end{equation}
where the sum runs over the $n$ phase-bins on the given NICER dataset; $\varpi = n - k$ are the estimated degrees of freedom, with $k = \# \{ \mathcal{C}, \chi, \xi, b, \kappa \bar{\gamma}, \varphi_o , \lambda_o \} = 7$ being the number of parameters of the model; and with $\varepsilon_{iN}$ being the statistical errors on the flux for each bin, as published in the cited NICER papers.

We use an L-BFGS-B algorithm from the Python \textit{SciPy optimization libraries} \cite{2020SciPy} to minimize $\chi_{r}^2$ over $\{ \lambda_o, \varphi_o, b \}$ for each configuration of the discrete sample in $(\mathcal{C}, \chi, \xi)$, restricting the beaming parameter to the range $b\in [0.5, 1.1]$. Finally, we refine the search in $\chi$  and $\xi$ 
simply by taking smaller intervals ($\Delta \chi=5 \degree$ and $\Delta \xi=1\degree$, respectively) at the most promising regions of the parameter space with the smallest \textit{reduced chi-square} values.

By applying this procedure to each of the four target pulsars, we  obtain combined approximate best-fits for their light curves and phase-averaged spectra (Sec.\ref{sec:LCandS}).


\section{Results} \label{sec:results}

We focus on a NS of radius $\sim$ $11.5\,$km that rotates with period $4.8\,$ms, which is close to the estimated periods of our four targets PSR J0437–4715, PSR J1231−1411, PSR J2124−3358 and PSR J0030+0451 (see Table \ref{table1})~\footnote{Recalling that, what is actually being fixed at the set of FF simulations, is the relation $R = 0.05 \, c \,  P/2\pi$. (As well as $\chi$ \& $\mathcal{C}$). }.

We first show the emission regions (ERs) arising from the GRFF simulations, defined as the regions at the stellar surface for which the four-current is spacelike (see Sec.~\ref{sec:emission_model}). In particular, we discuss the existence of an additional ER within the closed-zone of the pulsar, i.e., outside of the traditional polar caps along open field lines. 
Then, we present our simultaneous best-fits for the light curves and spectra of each target pulsar, following the methodology described in Sec.~\ref{sec:setup}. Part of the procedure is illustrated at Appx.~\ref{sec:appendix}, where the posterior probability density distributions are also shown. Finally, we analyze the relevance of these extra ERs on fitting the light curves and discuss the main limitations of the adopted emission model, specially regarding the flux normalization.

\begin{table}
\caption{ Relevant physical parameters (priors) of our four target millisecond pulsars. The third column is the magnetic field at the polar surface assuming the standard spin-down formula for a dipole in vacuum. Data are taken from the Australia Telescope National Facility pulsar catalogs \href{http://www.atnf.csiro.au/people/pulsar/psrcat/}{http://www.atnf.csiro.au/people/pulsar/psrcat/}.}
\vspace{0.2cm}
\centering {
\begin{tabular}{ c | c  c  c }         \hline \hline
~ PSR ~    & ~ period [ms] ~ & ~ $B_{\rm pole}$ [$10^8$ G] ~ & ~ distance [pc] ~ \\ \hline
J0437–4715 &  $~ 5.75 ~$  & $~ 5.80 ~$  & $~ 157 ~$ \\
J1231−1411 &  $~ 3.68 ~$  & $~ 2.93 ~$  & $~ 420 ~$ \\
J2124−3358 &  $~ 4.93 ~$  & $~ 3.22 ~$  & $~ 410 ~$ \\
J0030+0451 &  $~ 4.87 ~$  & $~ 2.25 ~$  & $~ 324 ~$ \\ \hline  \hline
\end{tabular}
}
\label{table1}
\end{table}

\subsection{Emitting regions} \label{sec:ERs}

\begin{figure*}
\begin{center}
\includegraphics[scale=0.18]{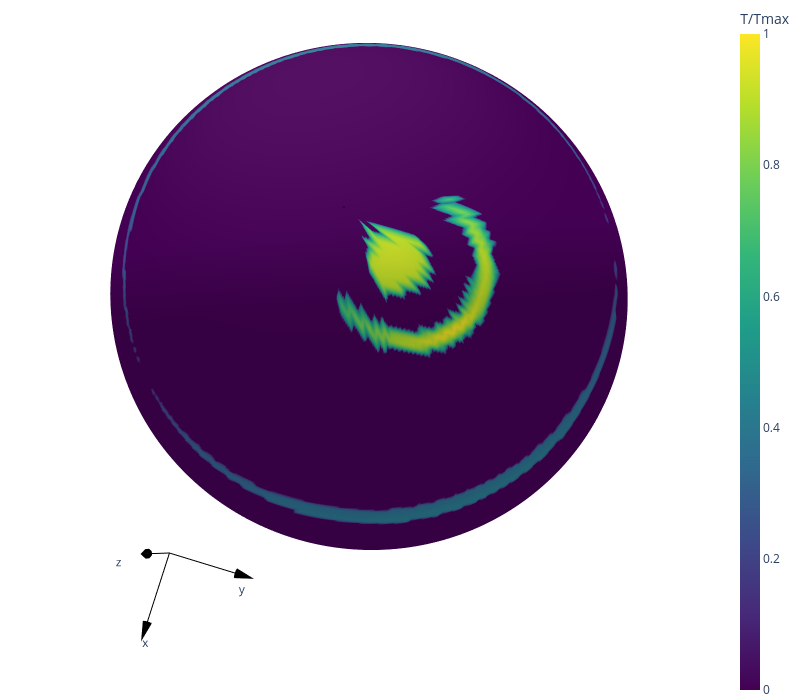}
\includegraphics[scale=0.18]{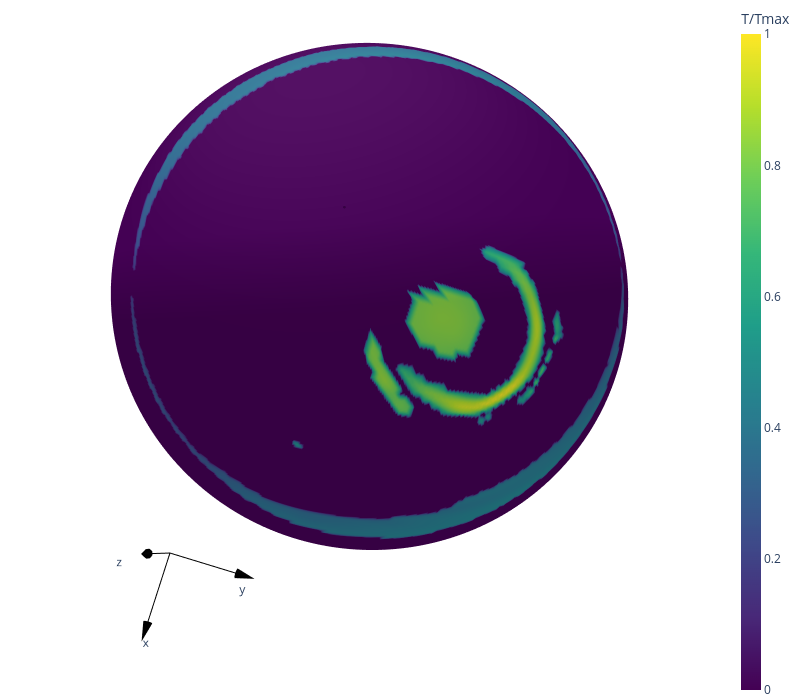}
\includegraphics[scale=0.18]{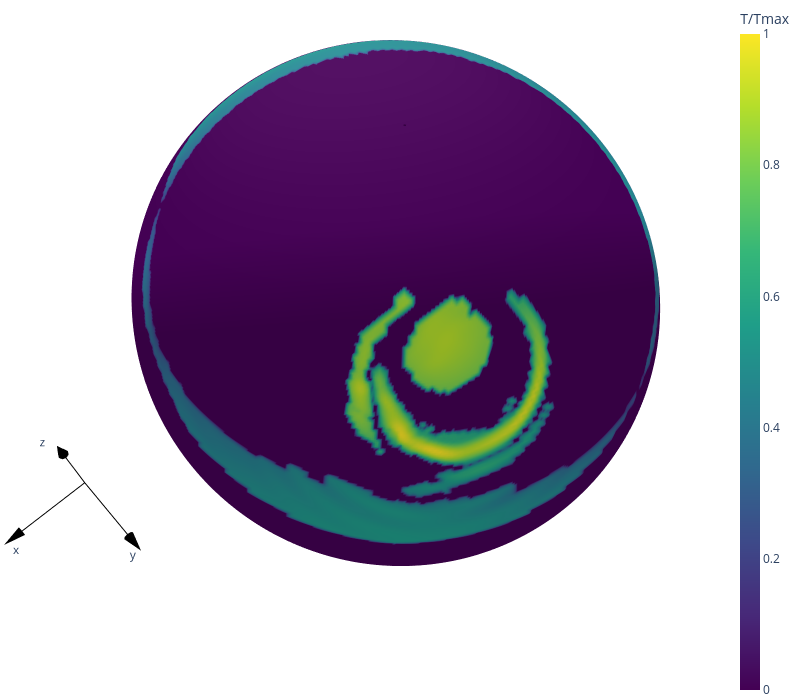}
\\
\vspace{0.4cm}
\includegraphics[scale=0.18]{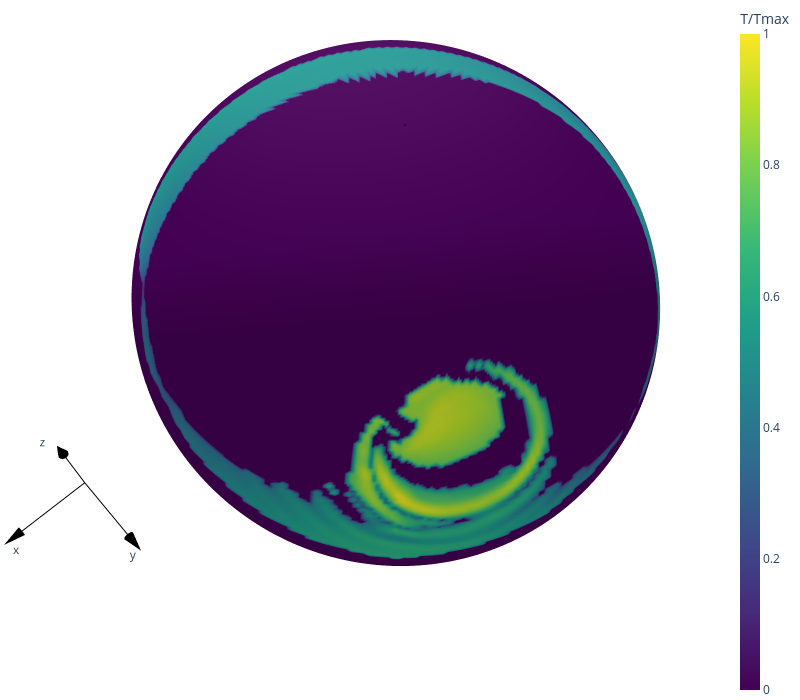}
\includegraphics[scale=0.18]{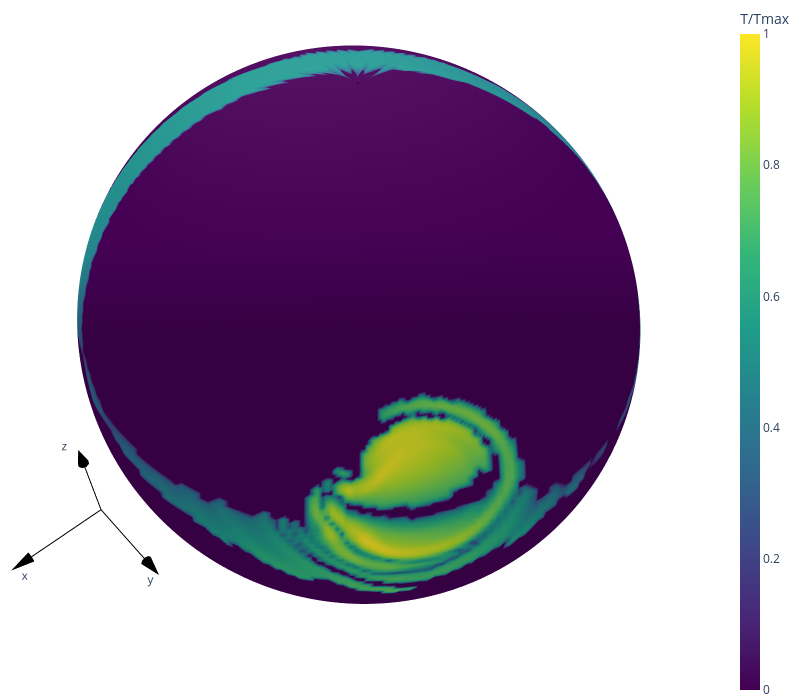}
\includegraphics[scale=0.18]{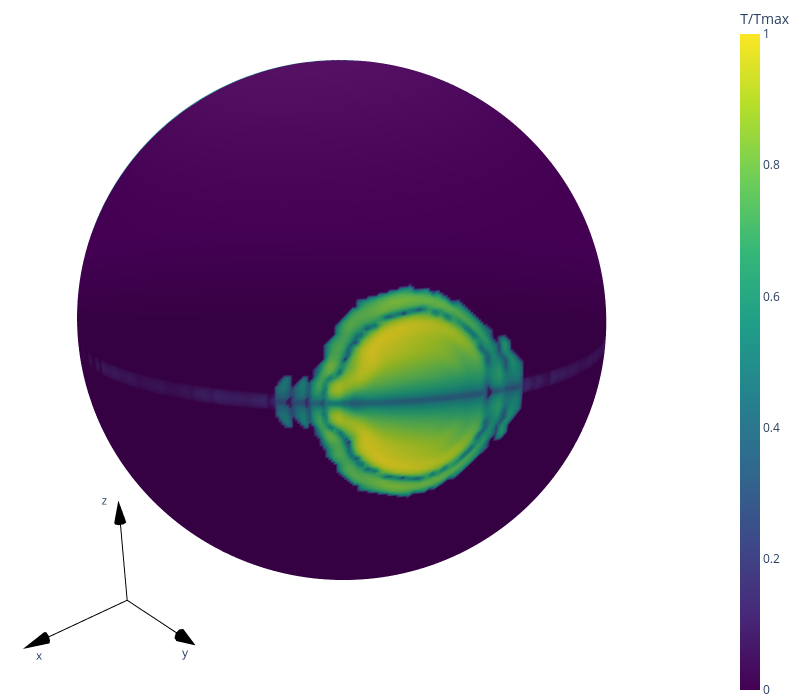}
\end{center}
\caption{ Emission regions for a pulsar with period $P\sim4.8\,$ms and compactness $\mathcal{C}=0.2$, for several misalignment angles $\chi= \{ 15\degree, 30\degree, 45\degree, 60\degree, 75\degree, 90\degree \}$ (from left-to-right and top-to-bottom). The colormaps show the temperature distribution at the NS surface (normalized by its maximum value $T_{max}$). That is, regions of spacelike current with temperature given by eq.~\eqref{eq:Teff}.
}
\label{fig:ER-chi}
\end{figure*}

\begin{figure*}
\begin{center}
\includegraphics[scale=0.18]{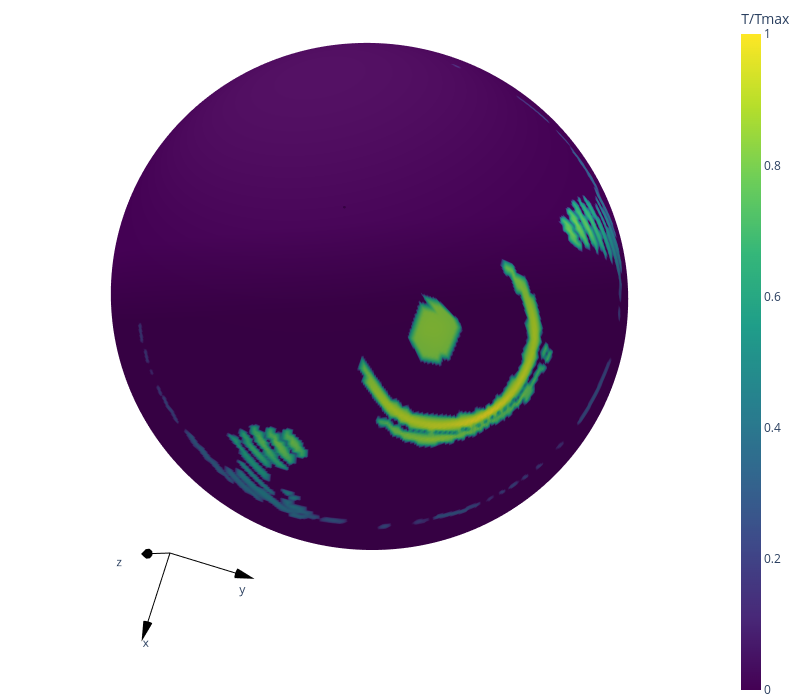}
\includegraphics[scale=0.13]{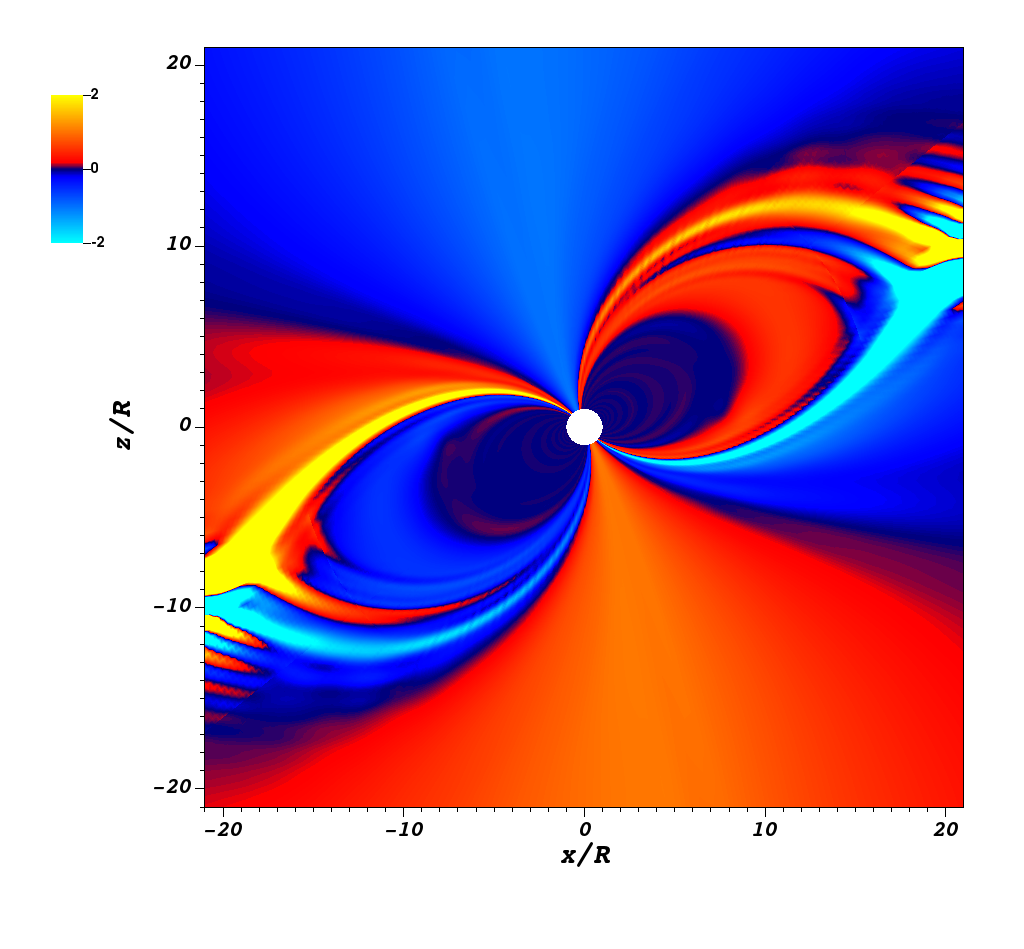}
\includegraphics[scale=0.13]{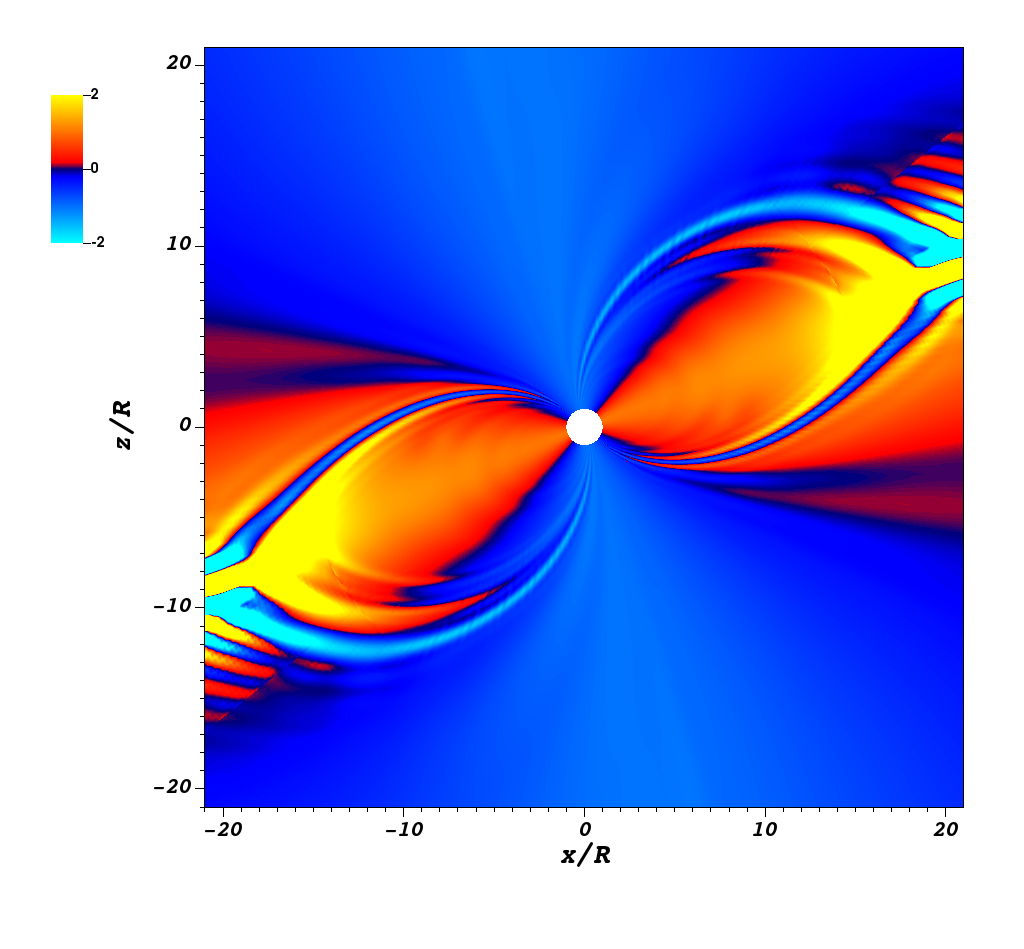}
\\
\vspace{0.4cm}
\includegraphics[scale=0.18]{cap_kerr01_C02_chi30.png}
\includegraphics[scale=0.13]{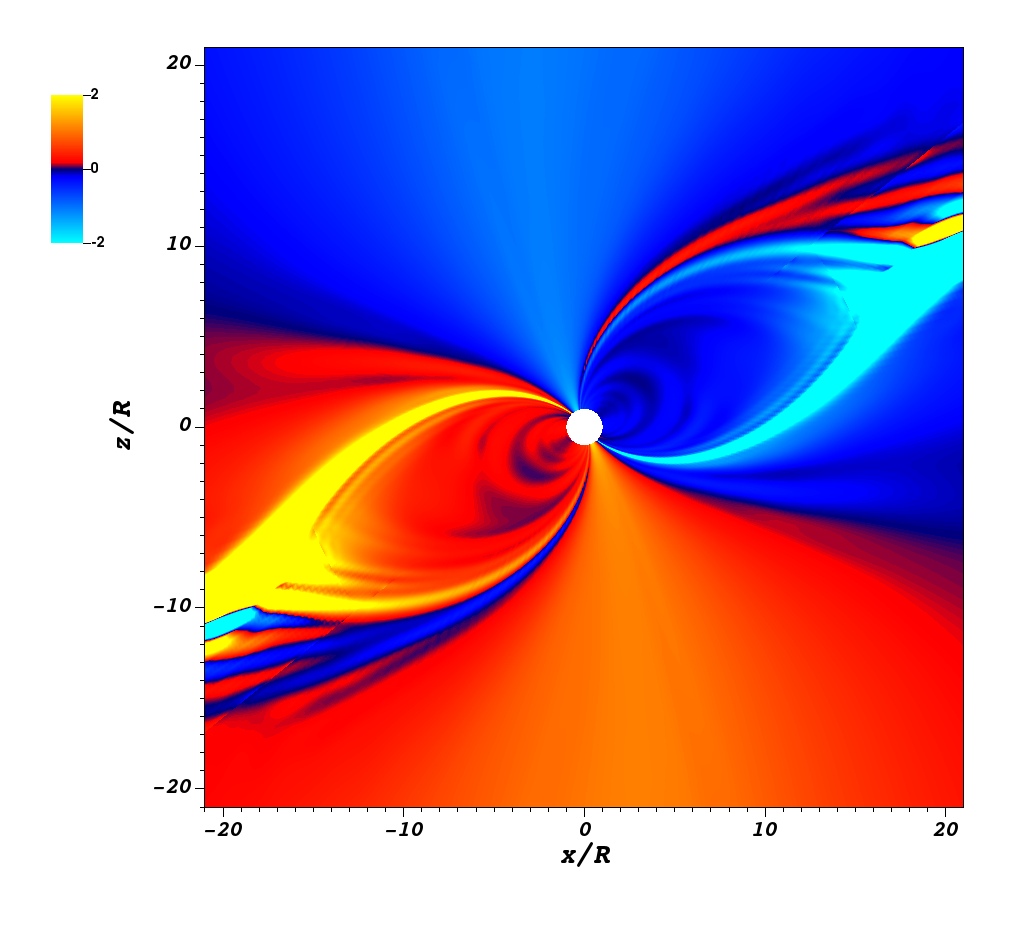}
\includegraphics[scale=0.13]{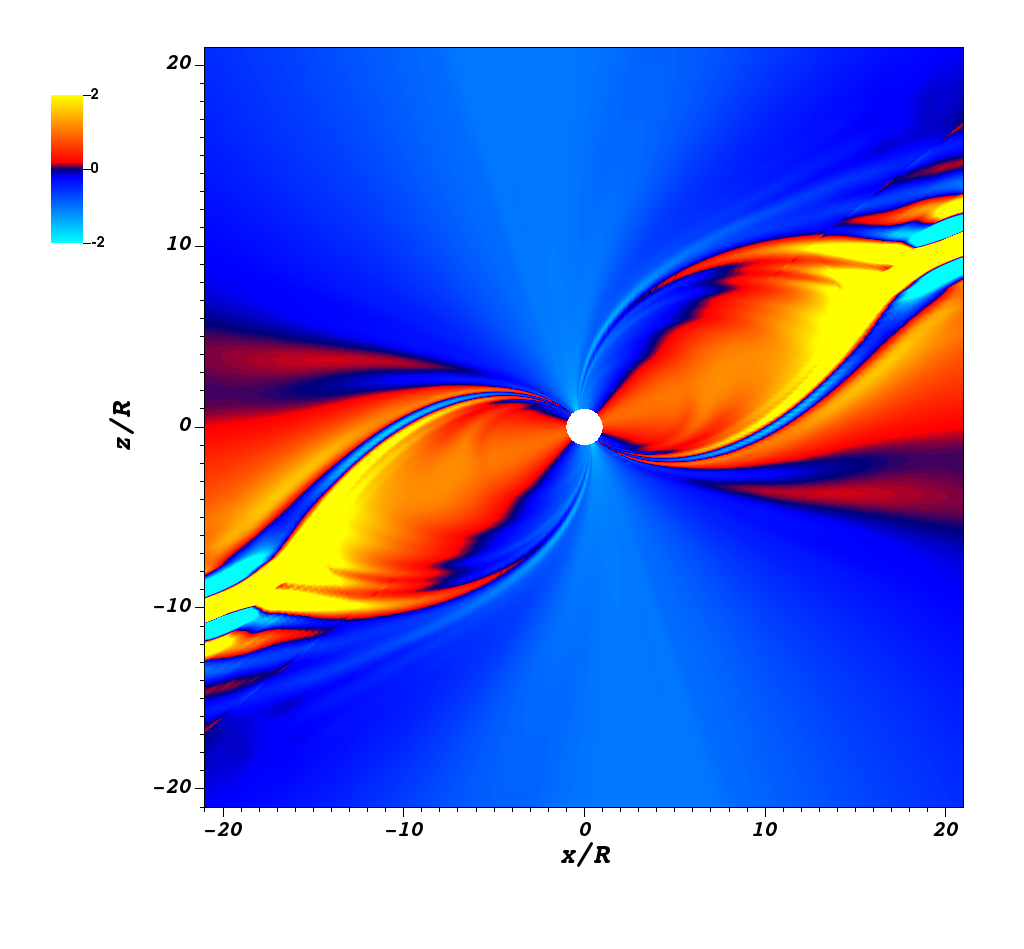}
\end{center}
\caption{ \textit{Emission region analysis for pulsar with period $P\sim4.8\,$ms and misalignment angle $\chi =  30\degree$}. FF configurations in flat (top panel)  
and Kerr (bottom panel) spacetimes. The colormap shows the 
temperature map for spacelike currents at the stellar surface (left); electric current along the magnetic field $J_{\parallel}/ \Omega B $ (middle) and charge distribution $\rho$ (right) on the meridional plane. The main difference is the presence of electric currents at the inner region of the pulsar closed-zone in curved spacetime.
}
\label{fig:ER-metric}
\end{figure*}

In Fig.~\ref{fig:ER-chi} we display the regions of spacelike four-current over the stellar surface of a NS with intermediate compactness $\mathcal{C}=0.2$, for different misalignments $\chi$. The 3D plots illustrates one hemisphere of the NS  showing (typically) a first ER, being a subset of the traditional polar caps (i.e. the regions demarcated by the footprints of all the open magnetic field lines). Interestingly, a second ER lies within the closed-zone of the pulsar, coincident with the intersection of the NS and the null charge surface (e.g., \cite{hirotani2001}). 
We analyze the origin of such additional spacelike currents by comparing these configurations to their flat spacetime analogues (i.e., coming from FF simulations with Minkowski metric).  In Figure \ref{fig:ER-metric}, we display the resulting ERs, together with the distribution of parallel current and electric charge densities at the meridional plane, both in flat and curved spacetimes. 

The comparison indicates that these additional spacelike currents arise from the existence of electric currents along closed field lines for GR pulsar magnetospheres (which are absent in flat spacetime), combined with the well known null charge surface that is present in both backgrounds. 
We checked that these electric currents are not just a transient effect, by extending one representative GRFF simulation to longer timescales. Additionally, we verified that they do not depend significantly on numerical resolution either. Our pulsar solutions are essentially converged at the resolution employed here, which can be further confirmed by looking at the luminosity within the light cylinder (as it was demonstrated in Fig.~2 of \cite{Pulsar}). 

If not stated otherwise, we include these extra "non-standard" ERs as part of our emission model. Moreover, we consider a single parameter $\kappa \bar{\gamma}$ to describe both emitting regions (i.e., over open/closed zones). With this choice, we would typically find that the open ER has an averaged effective temperature which is roughly twice the averaged temperature over the additional (closed) ER.

Since it seems plausible that the heating and emission from these two quite different regions could or should be modelled separately (see Sec.~\ref{sec:normalization} for more discussion),
we shall examine the possibility of decoupling the parameter $\kappa \bar{\gamma}$ for the particular case of PSR J0030+0451 (see Fig.\ref{fig:PSR0451-alt}). Additionally, we will also analyze what happens if we exclude these extra (closed) ERs, later in Sec.~\ref{sec:extraER}.

\subsection{Light curves and spectra} \label{sec:LCandS}

\begin{figure}
\begin{center}
\includegraphics[scale=0.55]{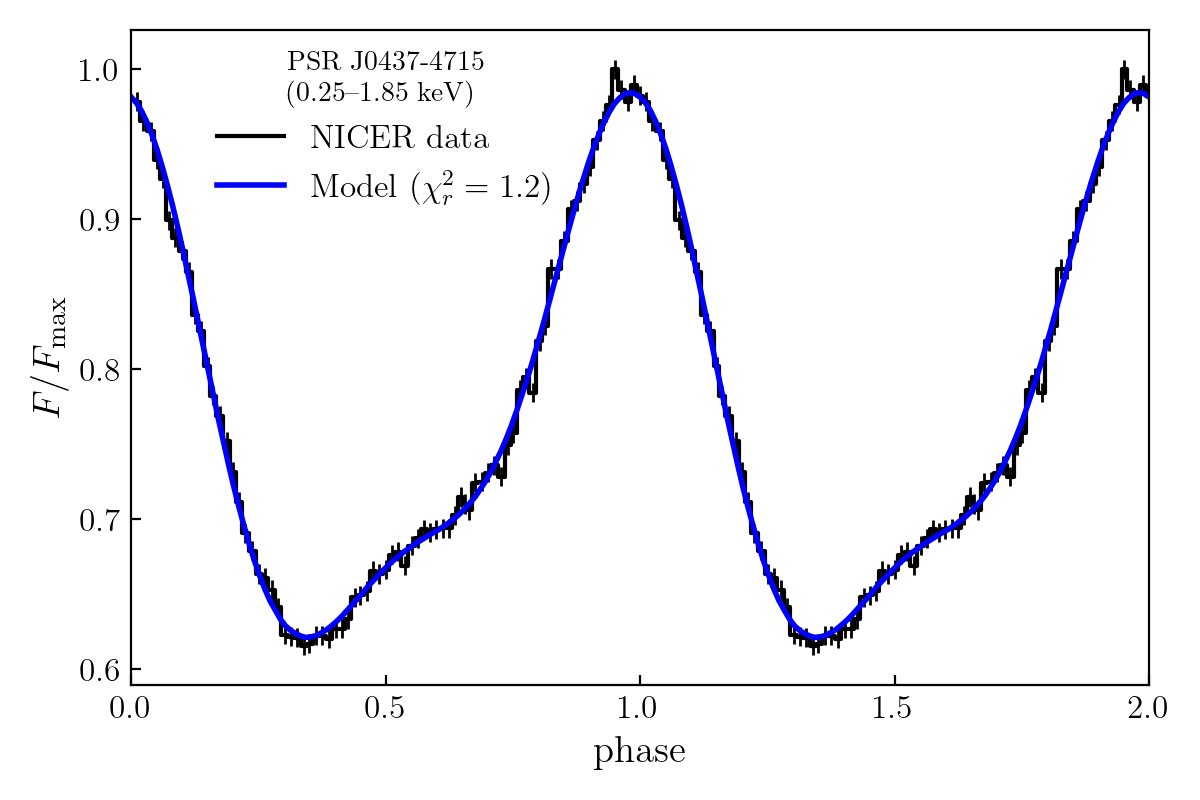}
\\
\includegraphics[scale=0.55]{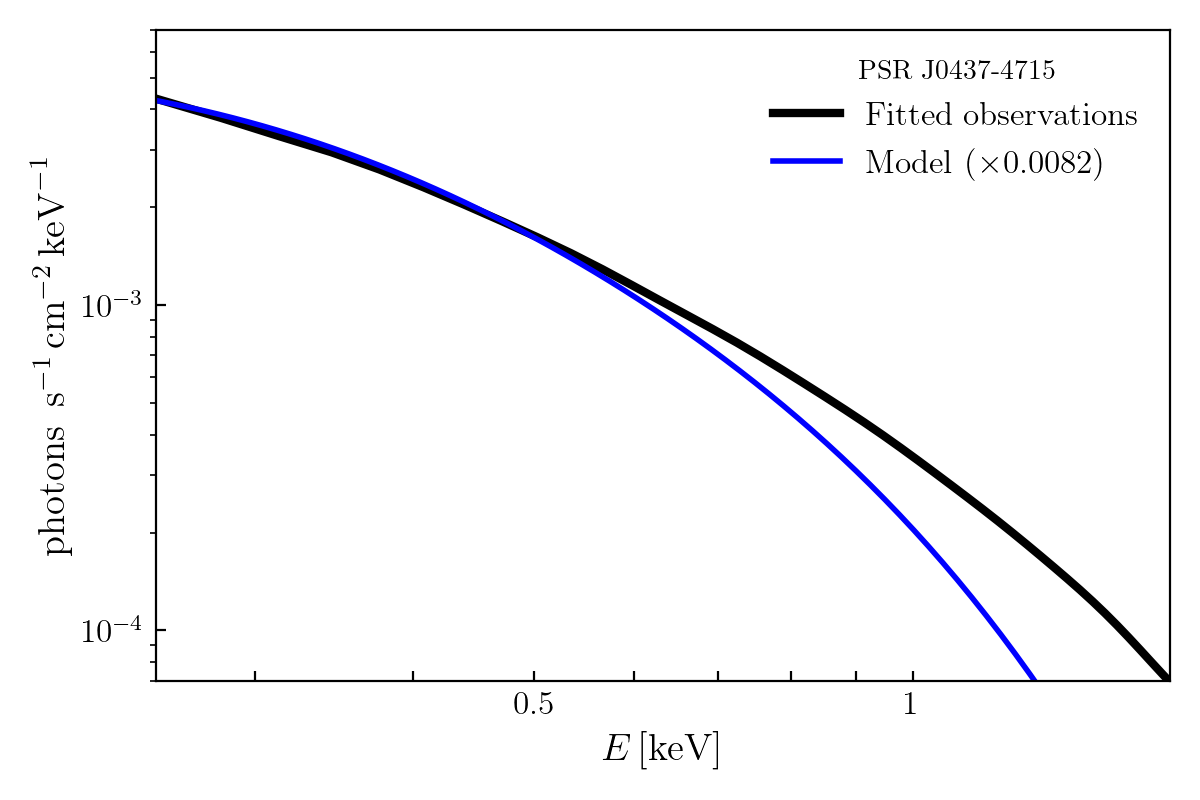}
\end{center}
\caption{ \textit{Best-fit to NICER X-ray data of PSR J0437-4715.}
Comparison of our best-fit configuration (first row in Table~\ref{table:best-fits}) against the observed light curve (top) and spectrum (bottom). The light curve shown in black, with its error bars, corresponds to NICER data taken from \cite{bogdanov2019constrainingI}.
Our modelled light curve (in blue) was integrated on the same energy window, as indicated in the plot.
The spectral data, on the other hand, is from XMM-Newton X-ray EPIC MOS1/2, fitted with a 2Hatm+BB+PL model \cite{bogdanov2012}, and we plotted (in black) the unabsorbed component. Our modelled spectrum (in blue) was re-scaled by a factor of $8.2 \times 10^{-3}$ for the comparison.  
}
\label{fig:PSR4715-best}
\end{figure}

\begin{figure}
\begin{center}
\includegraphics[scale=0.55]{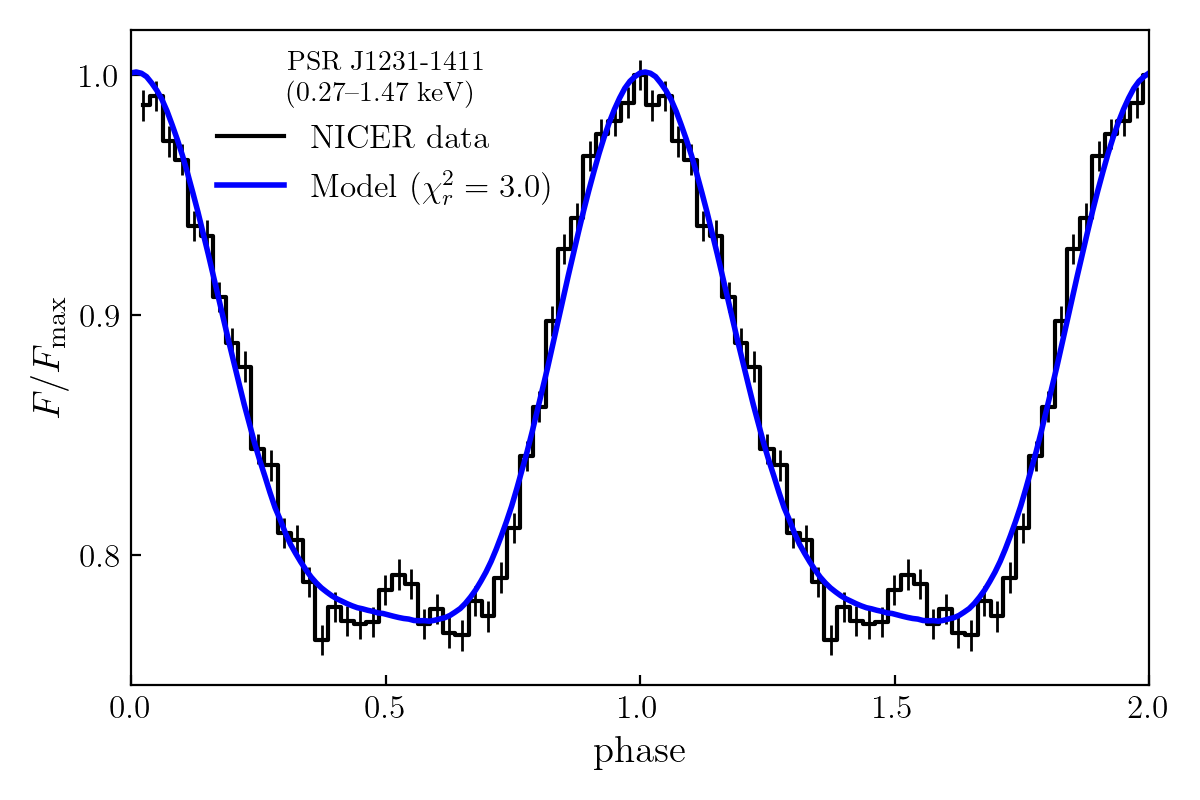}
\\
\includegraphics[scale=0.55]{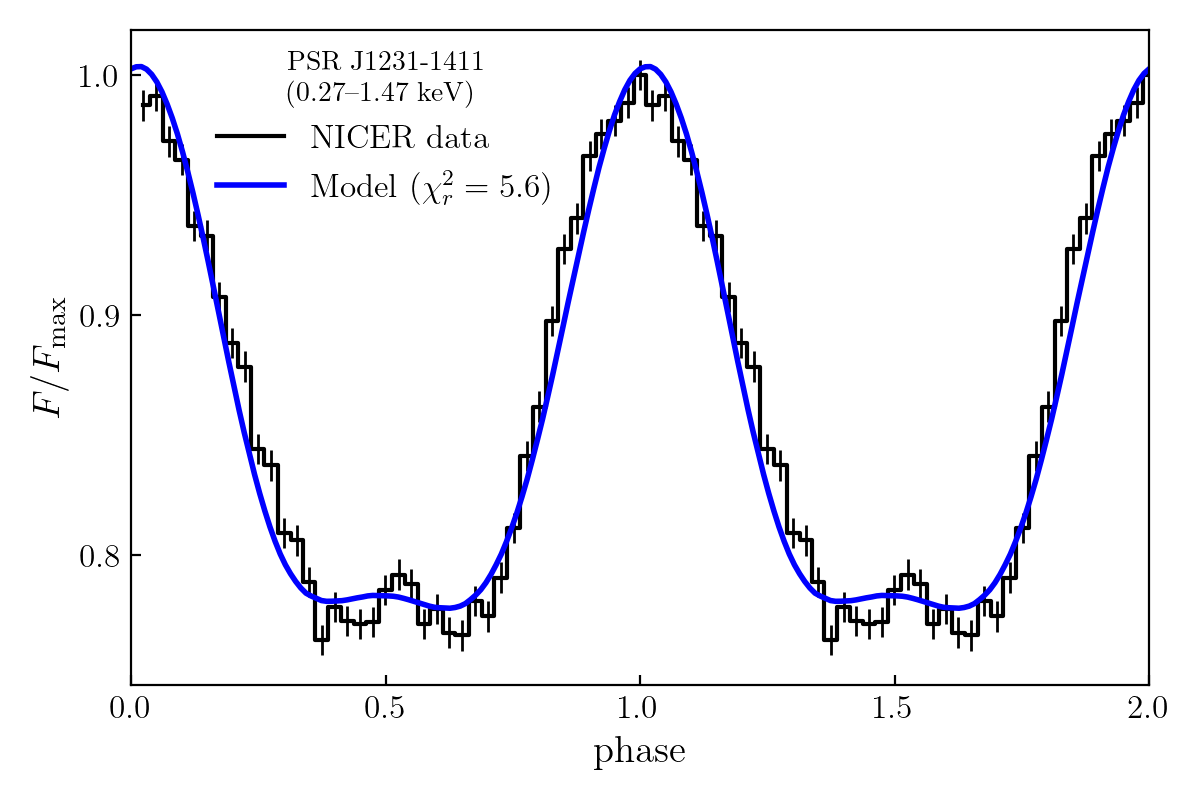}\\
\includegraphics[scale=0.55]{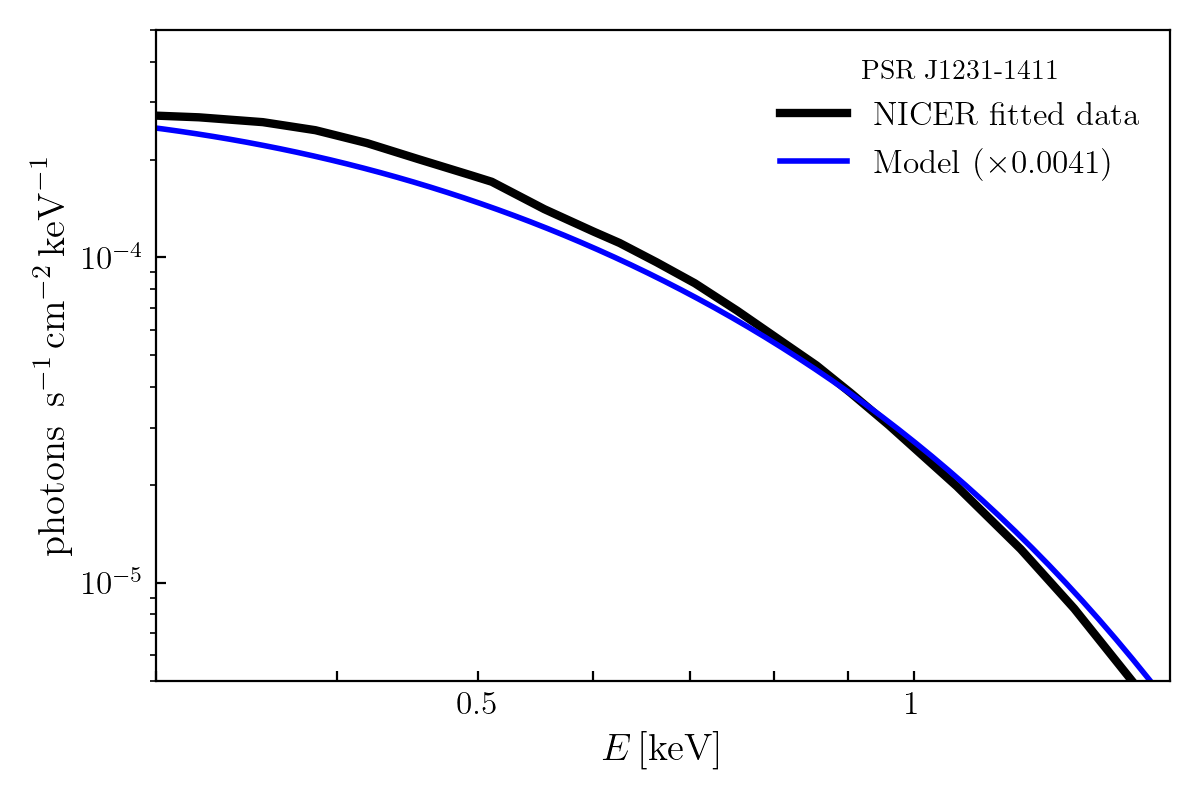}
\end{center}
\caption{ \textit{Best-fit to NICER X-ray data of PSR J1231-1411.} 
Comparison of our best-fit configuration (first row in Table~\ref{table:best-fits}) against the observed light curve (top) and spectrum (bottom). The light curve shown in black, with its error bars, correspond to NICER data taken from \cite{bogdanov2019constrainingI}.
Our modelled light curve (in blue) was integrated on the same energy window, as indicated in the plot.
The spectral data is from NICER, fitted with an Hatm (\textit{nsatmos}) model \cite{ray2019discovery}, and we plotted (in black) the unabsorbed component. Our modelled spectrum (in blue) was re-scaled by a factor of $4.1 \times 10^{-3}$ for the comparison. The middle panel corresponds to an alternative light curve fit (second row in Table~\ref{table:best-fits}) that better captures the qualitative features of the signal. 
}
\label{fig:PSR1411-best}
\end{figure}

\begin{figure}
\begin{center}
\includegraphics[scale=0.55]{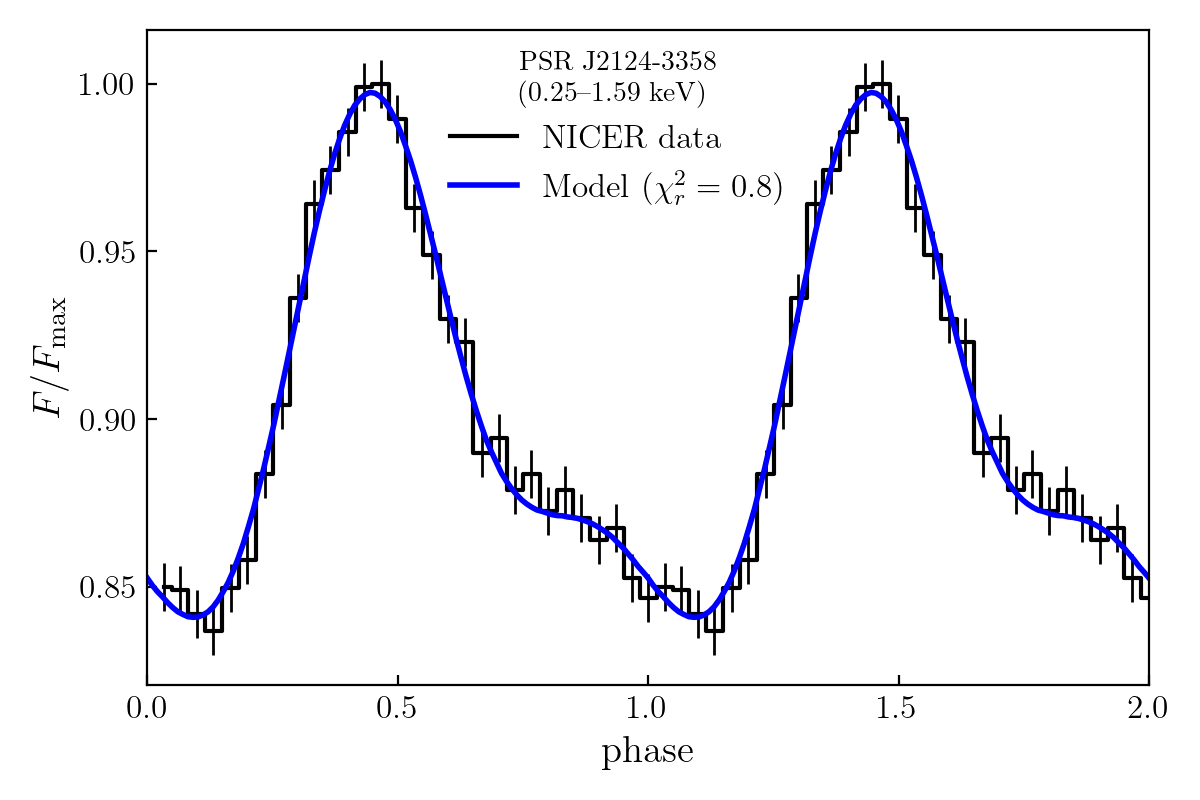}
\\
\includegraphics[scale=0.55]{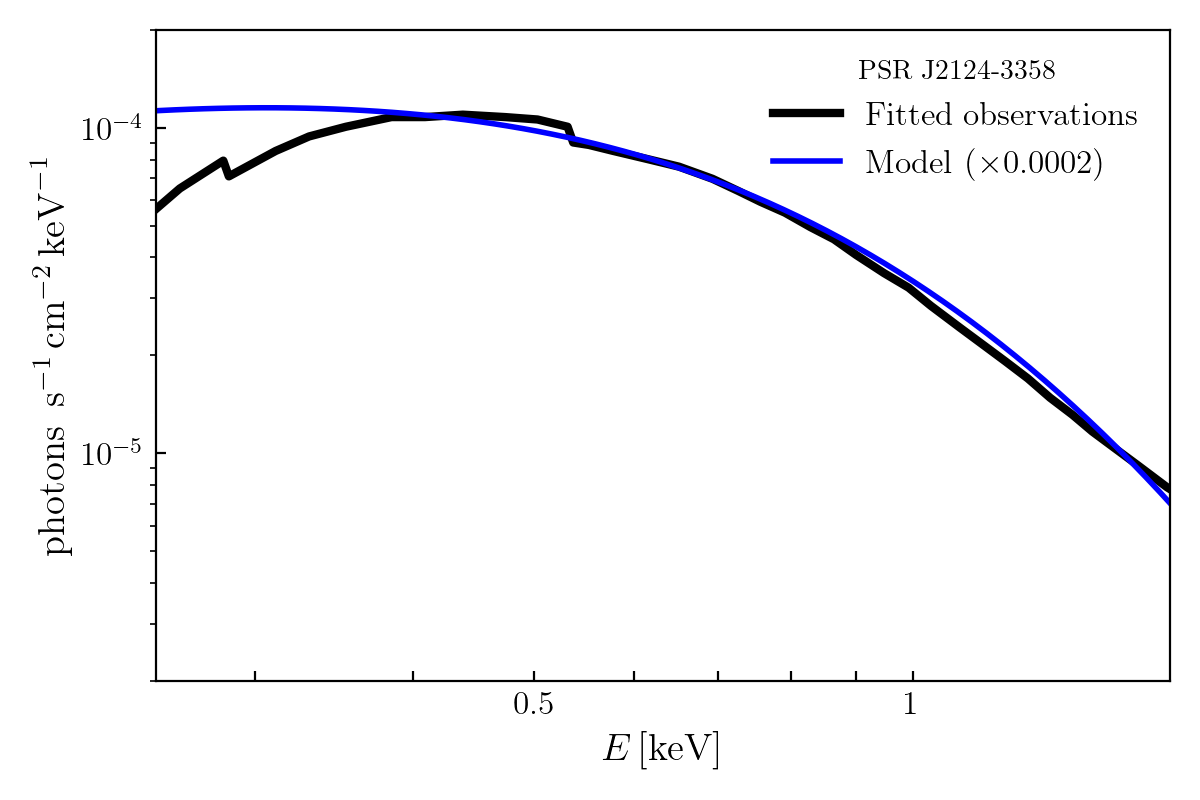}
\end{center}
\caption{ \textit{Best-fit to NICER X-ray data of PSR J2124-3358.} 
Comparison of our best-fit configuration (first row in Table~\ref{table:best-fits}) against the observed light curve (top) and spectrum (bottom). The light curve shown in black, with its error bars, correspond to NICER data taken from \cite{bogdanov2019constrainingI}.
Our modelled light curve (in blue) was integrated on the same energy window, as indicated in the plot.
The spectral data is taken from XMM-Newton observations, fitted with an Hatm+PL model \cite{zavlin2006xmm}, and we plotted (in black) the --only published-- absorbed component. Our modelled spectrum (in blue) was re-scaled by a factor of $2.1 \times 10^{-4}$ for the comparison.    
}
\label{fig:PSR3358-best}
\end{figure}

\begin{figure}
\begin{center}
\includegraphics[scale=0.55]{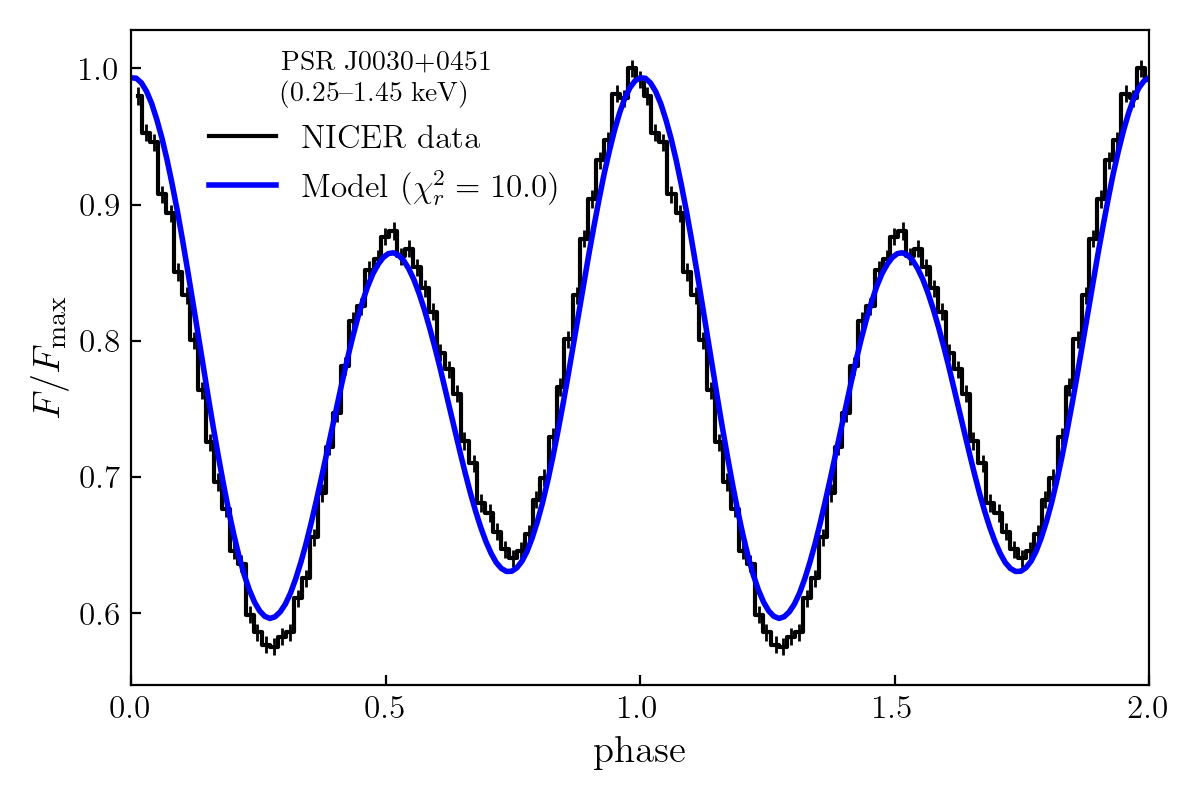}
\\
\includegraphics[scale=0.55]{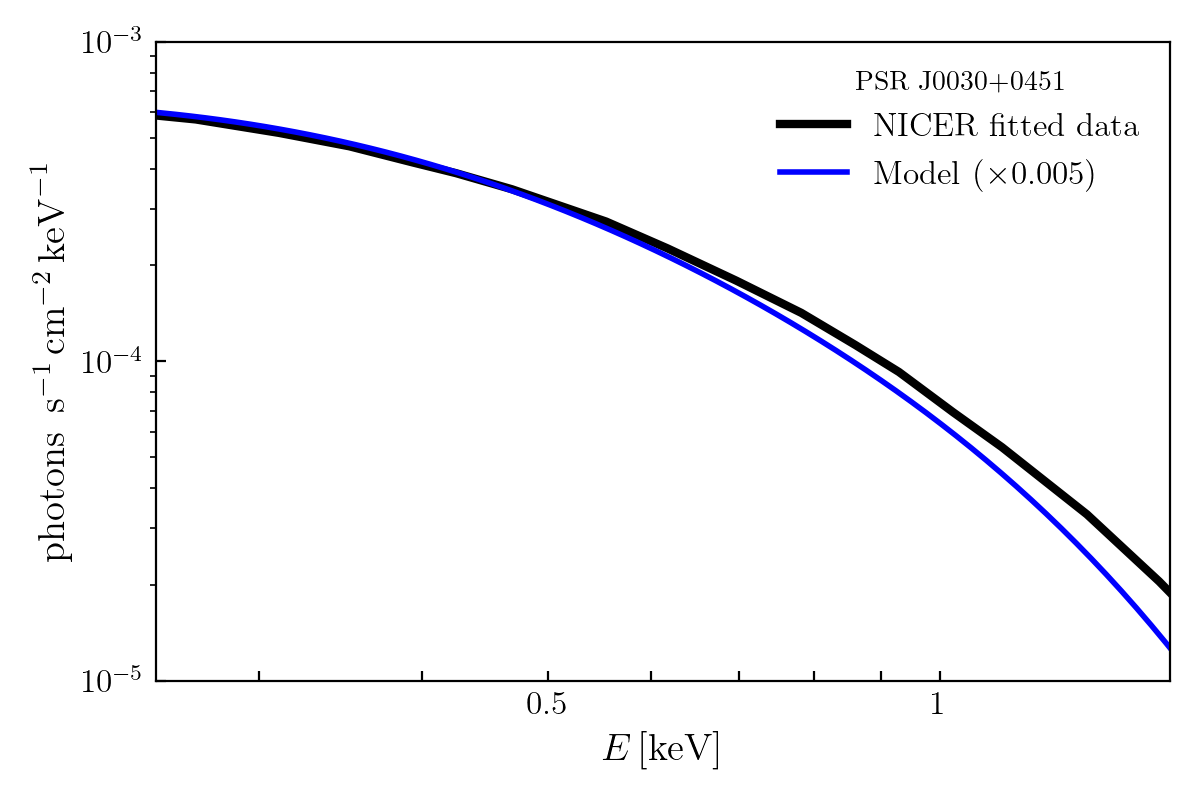}
\end{center}
\caption{ \textit{Best-fit to NICER X-ray data of PSR J0030+0451.} 
Comparison of our best-fit configuration (first row in Table~\ref{table:best-fits}) against the observed light curve (top) and spectrum (bottom). The light curve shown in black, with its error bars, correspond to NICER data taken from \cite{bogdanov2019constrainingI}.
Our modelled light curve (in blue) was integrated on the same energy window, as indicated in the plot.
The spectral data is taken from the unabsorbed posterior-expected phase-averaged spectrum for ST+PST (i.e., Hatm model) \cite{riley2019}, consistent with both XMM-Newton and NICER data.
Our modelled spectrum (in blue) was re-scaled by a factor of $5 \times 10^{-3}$ for the comparison.
}
\label{fig:PSR0451-best}
\end{figure}

\begin{figure}
\begin{center}
\includegraphics[scale=0.55]{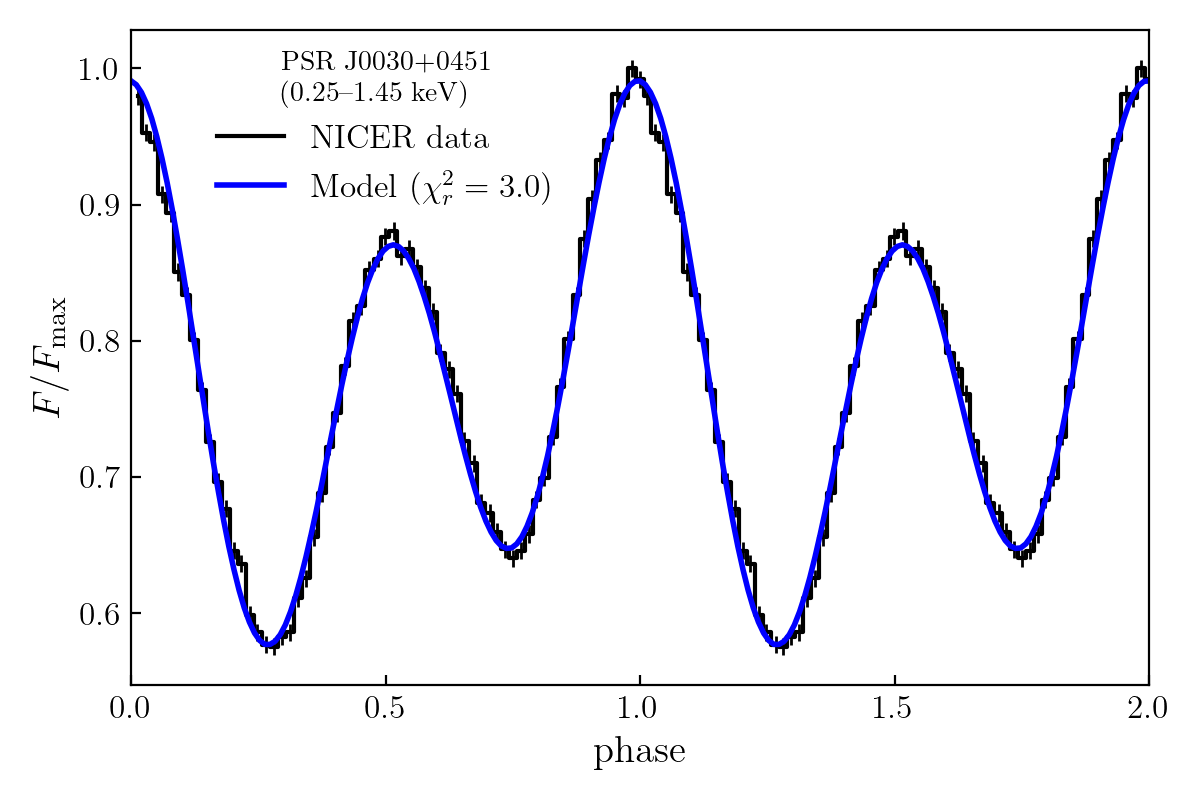}
\\
\includegraphics[scale=0.55]{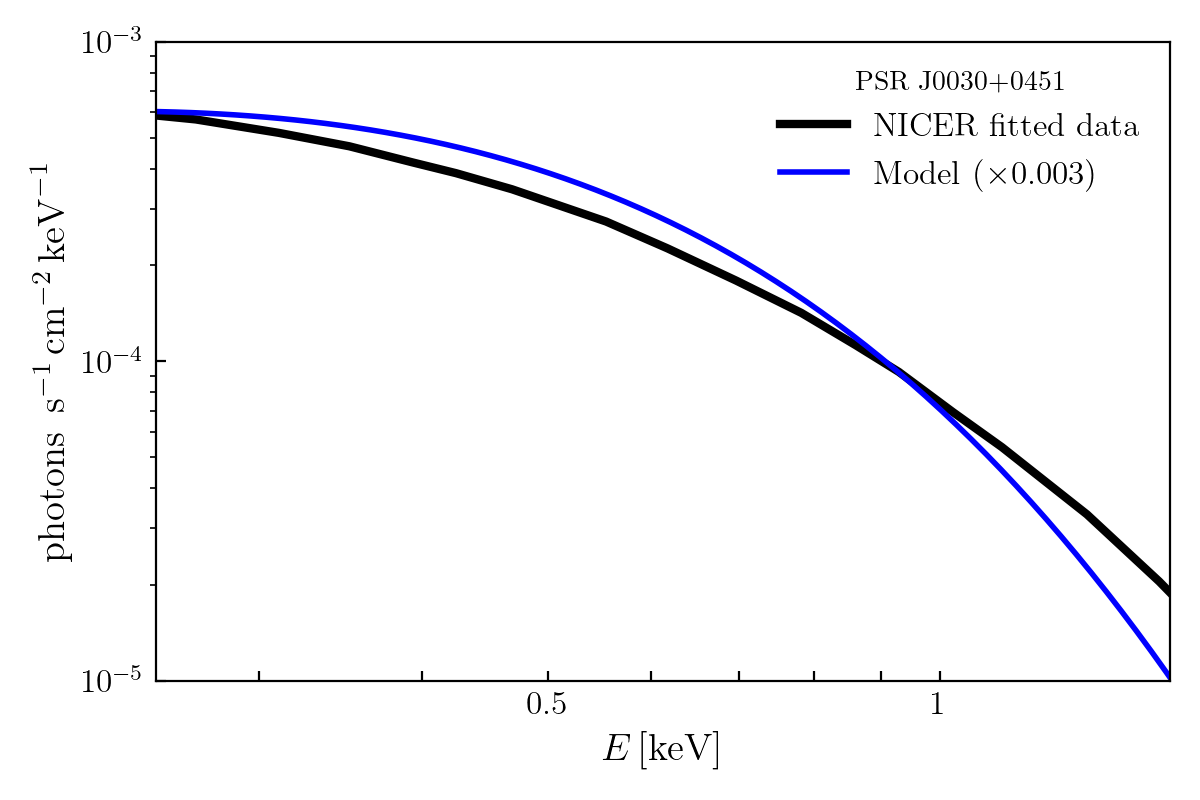}
\end{center}
\caption{ \textit{Alternative best-fit for PSR J0030+0451 light curve, by decoupling the temperature parameter $\kappa \bar{\gamma}$.} The effective temperature parameter $\kappa \bar{\gamma}$ is decoupled here for the two different kind of emission regions, i.e. over the polar cap (open field lines) and the extra ER over the closed zone of the pulsar. Being $(\kappa \bar{\gamma})_{\rm open} = 3.8 \times 10^7$ and $(\kappa \bar{\gamma})_{\rm closed} \approx 4 \, (\kappa \bar{\gamma})_{\rm open}$.
All the other specifications of the plots are identical to Fig.~\ref{fig:PSR0451-best}.
}
\label{fig:PSR0451-alt}
\end{figure}

We present our best-fits to the pulsar light curves and spectral distributions in Figures \ref{fig:PSR4715-best} to \ref{fig:PSR0451-alt}, and summarize relevant information about the parameters in Table \ref{table:best-fits}.
Even though the exploration of the parameter space is not fully exhaustive\footnote{Meaning that our coverage for some of the parameters, like the stellar compactness $\mathcal{C}$ and the pulsar misalignment $\chi$, is severely limited by the computational cost of the GRFF simulations.}, we achieve very good simultaneous fits to both the light curves and spectra for all the target pulsars.
In particular, for PSR J0437–4715 and PSR J2124−3358, we obtain really good light curve fits for which the \textit{reduced chi-square} factors are close to unity. 
Whereas for the other two, PSR J1231−1411 and PSR J0030+0451, we find minimum values around $\chi_{r}^2 \sim 3$.
It is worth emphasizing again that we are assuming a centered dipole magnetic field configuration (i.e., we do not include any centered nor off-centered higher multipoles). 
 Our model has only few parameters, and the observational X-ray data for these pulsars is very accurate. The values of $\chi_{r}^2$ attained here are, hence, quite encouraging.

All the light curve fittings are accompanied by a consistent phase-averaged spectral distribution on their relevant energy windows (bottom panels). The only mild exception being the spectrum of PSR J0437–4715, which presents a visible deviation at high energies. On the other hand, the reason for the apparent deviation for the PSR J1231−1411 spectrum at low energies in Fig.~\ref{fig:PSR1411-best} is simply that we are comparing 
it against the absorbed fitted observational data, an absorption that mainly affects at the lower energies. This fact imply a larger uncertainty in the determination of $\kappa \bar{\gamma}$ for this pulsar, although we notice that it does not affect significantly the light curve fitting; that is, for a wide range of $\kappa \bar{\gamma}$ values, we find best-fits with almost identical \textit{reduced chi-square} factors at the same values of $\mathcal{C}$, $\chi$ and $\xi$. 
Notice that we are comparing our spectra with respect to some given fitted observational data, taken from different papers. There is, thus, an inherent uncertainty in the data which is not indicated on the plots.

As apparent from Table \ref{table:best-fits}, there are small degeneracies on the best fits among the geometrical parameters of the system, like the angles $\chi$ and $\xi$ (e.g., in the case of PSR J0437–4715), or $\mathcal{C}$ (as the case of PSR J2124−3358, which presents various configurations with $\chi_{r}^2 \sim 1$). This kind of degeneracies are commonly encountered, 
and they could potentially be resolved by simultaneously contrasting (via an additional modelling) with electromagnetic observations in the radio and/or gamma-ray bands (e.g., \cite{chen2020, kalapotharakos2021}). 

In the light curve of PSR J1231−1411, there is a small (but clear) secondary peak in the signal, which turns out to be quite challenging to accurately capture. In particular, we find that the configurations with minimum values of $\chi_{r}^2$ fail to reproduce this feature of the signal. 
However, we do find other configurations with slightly larger $\chi_{r}^2$ values that provide a better qualitative fit (see middle panel of Fig.\ref{fig:PSR1411-best}).

The most challenging X-ray light curve is, as perhaps expected, that of PSR J0030+0451. For this case, 
we get a more modest fitting with a factor $\chi_{r}^2 = 10$ in the general search, although the result is still qualitatively good, in the sense that we are able to capture the significant difference in the amplitude of profiles and the depths of valleys. Moreover, we note that it is possible to significantly improve the light curve fitting (achieving $\chi_{r}^2 =3$) if allowing a decoupling of the effective temperature coefficient $\kappa \bar{\gamma}$ for the two kinds of ERs (i.e., over open/closed regions).
This is illustrated on Fig.~\ref{fig:PSR0451-alt}, where the quality of this alternative light curve fit can be fully appreciated. It is interesting to note how the inclusion of a single additional parameter in our model, in this case, can considerably decrease the \textit{reduced chi-square} factor.
In doing so, on the other hand, we observe there is a negative impact on its spectral distribution (see bottom panel of Fig.~\ref{fig:PSR0451-alt}).
We shall come back to this point below, when discussing the limitations of the emission model.

\begin{table}
\caption{ Best-fits parameters for the X-rays signals of the four target millisecond pulsars. We fixed a single approximate rotation period of $4.8\,$ms for all cases, while taking individual priors for magnetic field strength and distance from Table~\ref{table1}.\\
{\footnotesize (*) The effective temperature parameter $\kappa \bar{\gamma}$ is decoupled here for the two different kind of emission regions, i.e. over the polar cap (open field lines) and the extra ER over the closed zone of the pulsar. Being $(\kappa \bar{\gamma})_{\rm open} = 3.8 \times 10^7$ and $(\kappa \bar{\gamma})_{\rm closed} \approx 4 \, (\kappa \bar{\gamma})_{\rm open}$.}
}
\vspace{0.2cm}
\centering {
\begin{tabular}{ c | c  c  c    c   c  | c}         \hline \hline
~ PSR ~    & ~ $\mathcal{C}$ ~ & ~ $\chi$ ~ & ~ $\xi$ ~  & ~ $b$ ~ & ~ $\kappa \bar{\gamma}$ ~ & $\chi_{r}^2$\\ \hline
J0437–4715 &  $~ 0.25 ~$  & $~ 20 \degree ~$  & $~ 53 \degree ~$ & $~ 0.94 ~$  & $~ 7.0 \times 10^6 ~$ & $~ ~ 1.2~$ \\ 
&  $~ 0.25 ~$  & $~ 15 \degree ~$  & $~ 61 \degree ~$ & $~ 1.00 ~$  & $~ 7.0 \times 10^6 ~$ & $~ ~ 1.2~$ \\
J1231−1411 &  $~ 0.18 ~$  & $~ 75 \degree ~$  & $~ 12 \degree ~$ & $~ 0.81 ~$  & $~ 2.5 \times 10^7 ~$ & $~ ~ 3.0~$ \\
           &  $~ 0.22 ~$  & $~ 43 \degree ~$  & $~ 30 \degree ~$ & $~ 0.60 ~$  & $~ 2.5 \times 10^7 ~$ & $~ ~ 5.3~$ \\
J2124−3358 &  $~ 0.22 ~$  & $~ 60 \degree ~$  & $~ 24 \degree ~$ & $~ 0.67 ~$  & $~ 1.3 \times 10^8 ~$ & $~ ~ 0.8~$ \\
&  $~ 0.20 ~$  & $~ 75 \degree ~$  & $~ 17 \degree ~$ & $~ 0.67 ~$  & $~ 1.3 \times 10^8 ~$ & $~ ~ 0.9~$ \\
J0030+0451 &  $~ 0.22 ~$  & $~ 25 \degree ~$  & $~ 85 \degree ~$ & $~ 1.09 ~$  & $~ 3.8 \times 10^7 ~$ & $~ 10.0~$ \\
&  $~ 0.22 ~$  & $~ 25 \degree ~$  & $~ 80 \degree ~$ & $~ 1.02 ~$  & (*) & $~ 3.0~$ \\ \hline  \hline
\end{tabular}
}
\label{table:best-fits}
\end{table}

\subsection{Relevance of the additional emitting region} \label{sec:extraER}

\begin{figure*}
    \subfloat[ PSR J0437-4715 \label{split:a}]{%
      \includegraphics[scale=0.56]{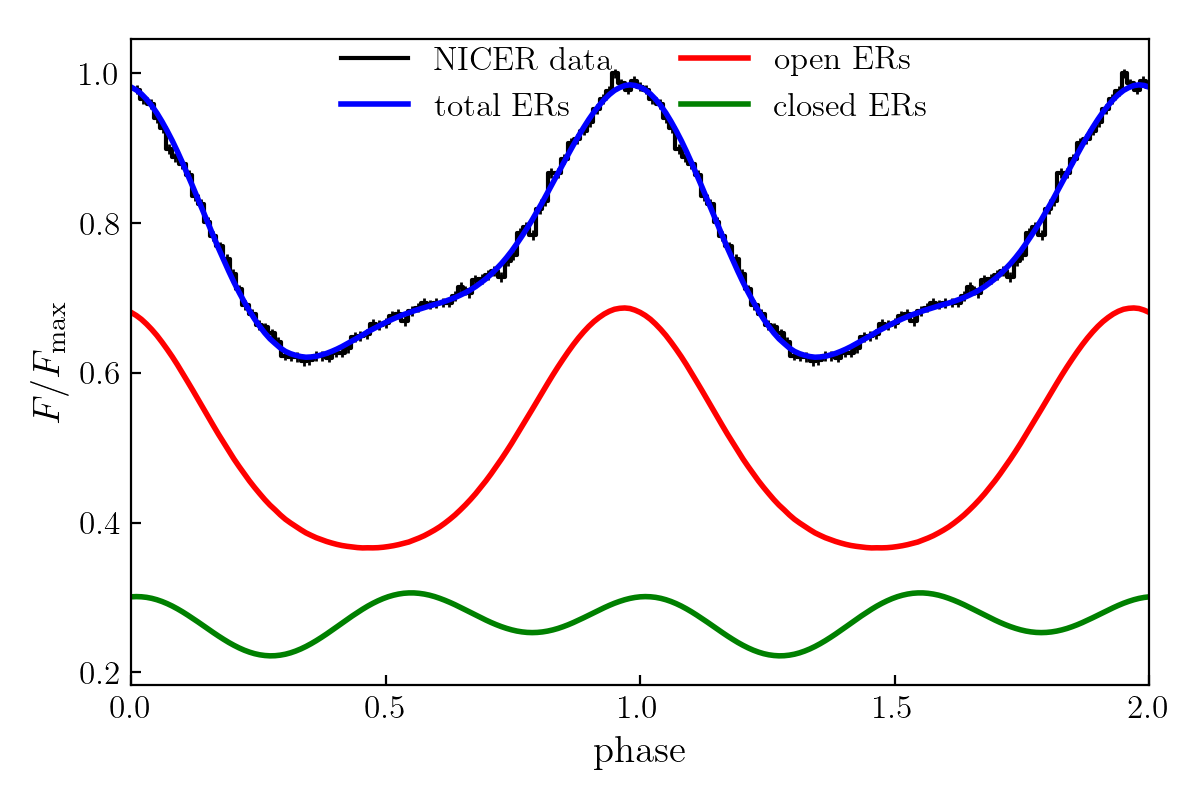}    
    }
    \subfloat[PSR J1231-1411 \label{split:b}]{%
      \includegraphics[scale=0.56]{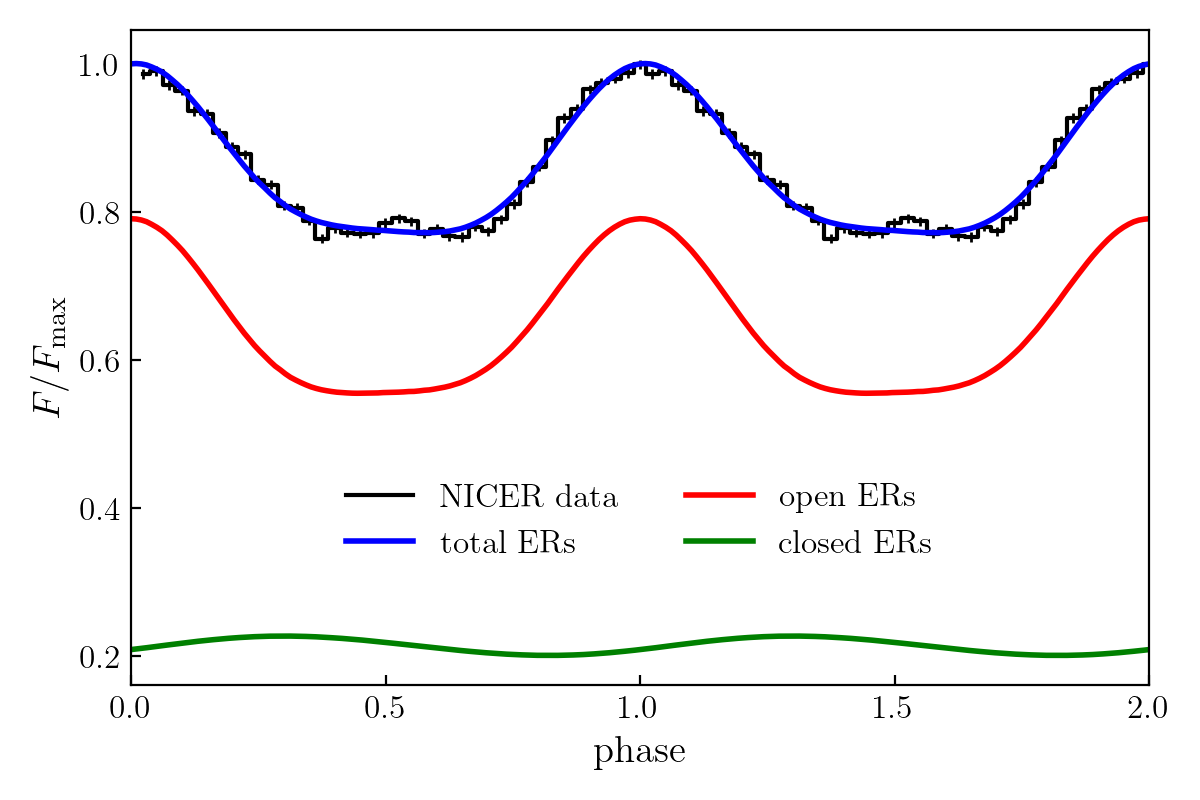}
    }
    \vspace{0.25cm}
    \subfloat[PSR J2124-3358 \label{split:c}]{%
      \includegraphics[scale=0.56]{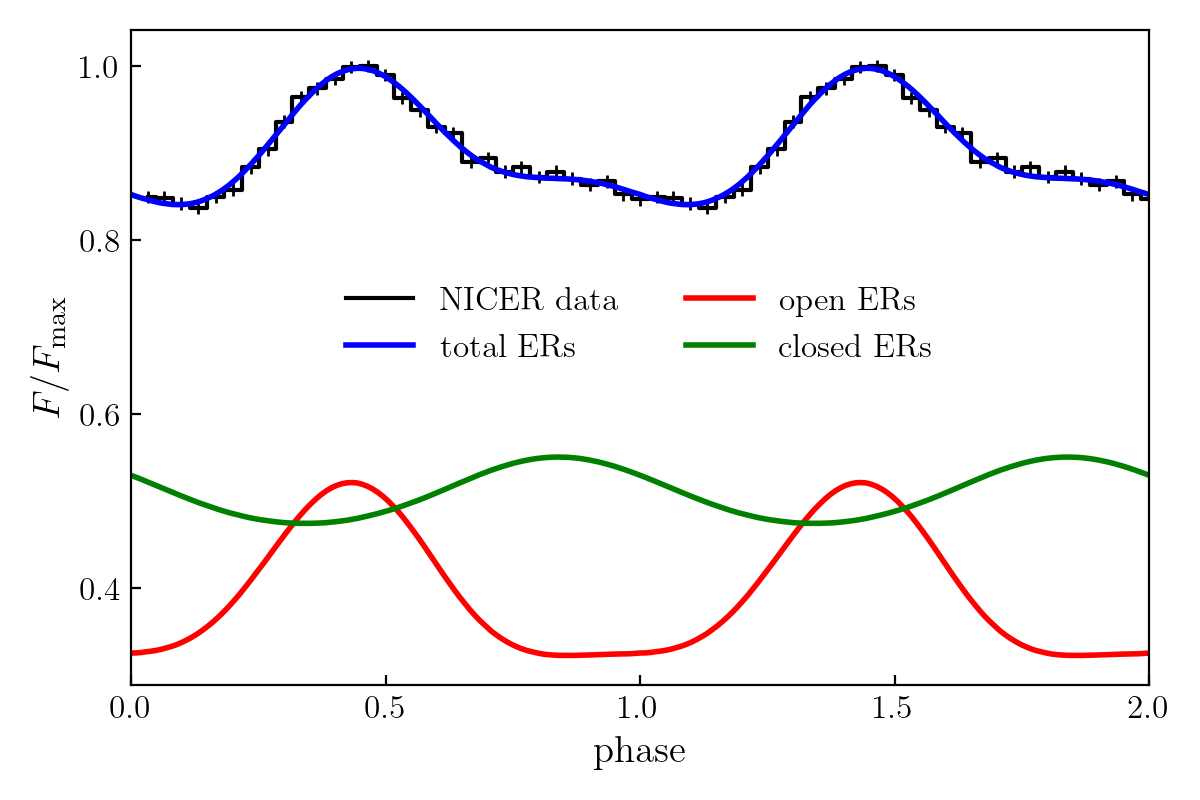}
    }
    \subfloat[PSR J0030+0451 \label{split:d}]{%
      \includegraphics[scale=0.56]{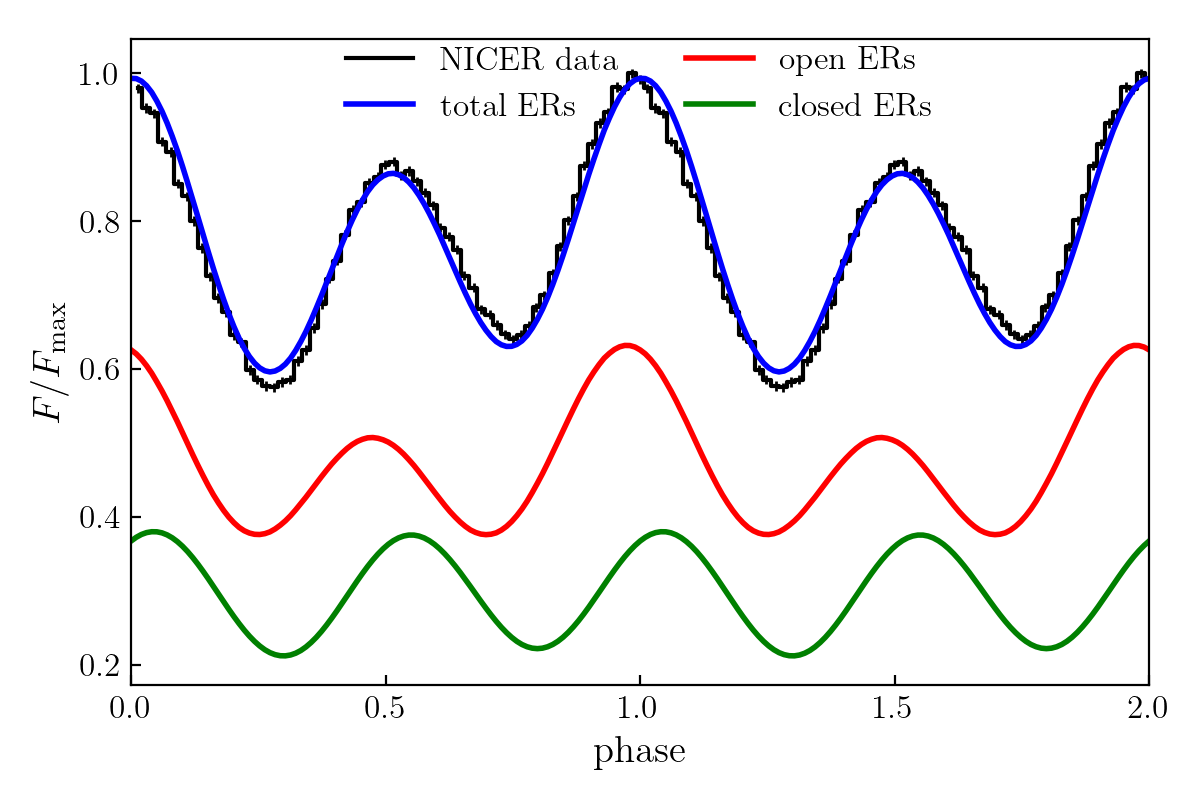}
    }
    \vspace{0.5cm}    
    \caption{\textit{Contribution of the non-standard (closed) ERs to the light curves for each of the target pulsars (a), (b), (c) and (d).} We split the total ERs (i.e., spacelike currents) on a piece contained within the usual polar caps (i.e., open field lines) region, and an extra piece arising from a spacetime curvature effect at the closed-zone of the pulsar. The plots correspond to the first configurations listed in Table~\ref{table:best-fits} for each of the four target pulsars.
    }
    \label{fig:LC-extra}
  \end{figure*}

We want to elucidate now the importance of the non-standard ER found over the closed-zone of the pulsar. To this end, we first examine the impact of such additional ER on the light curves best-fits from the previous section. In Figure \ref{fig:LC-extra} we present a splitting of each light curve into the contributions arising from the open/closed ERs. 
It can be seen that the signals from the open ER, sub-regions of the antipodal polar caps, tend to be dominant and highly symmetric. While those arising from the ER in the closed-zone, also antipodal and quite symmetric, are slightly out of phase with respect to the open ER curve. This is what allows to produce a combined signal with an asymmetric interpulse between the peaks as in PSR J0437–4715 and PSR J2124−3358. Or, in the case of PSR J0030+0451, to approximate the observed imbalance between the two valleys on its light curve.

However, one might ask what happens if the closed ER is excluded, while allowing for a full exploration of the parameter space. Is it possible to achieve fits equally good as those of Table \ref{table:best-fits} where both ERs were included? 

The answer is generally NO, with the exception of the PSR J1231−1411 signal. We filtered-out the closed ER and repeated the whole search procedure of Sec.\ref{sec:search} for all our target pulsars. The results of such exploration are summarized on Table \ref{table:filt-fits}, where one clearly sees worse quantitative fits, reflected on $\chi_{r}^2$ values increasing by a factor up to $\sim 8$. The reason is, as anticipated above, that without the additional (closed) ER, the model is unable to capture the observed features between the main pulses in the signals. 
In PSR J1231−1411, such interpulse is quite small and symmetric, explaining why this case can be fitted almost equally well with both (open/closed) and a single (open) emitting regions. This is consistent with previous results by other collaborations for PSR J0030+0451, according to which a single pair of antipodal regions is unable to properly capture the asymmetries of the pulse profiles \cite{miller2019, riley2019, chen2020, kalapotharakos2021}. The previous works resolve the tension by introducing off-centered higher magnetic multipoles, thus giving place to more complex polar caps. In contrast, in our case the tension is resolved by having another pair of antipodal regions (the closed ERs, arising from the GRFF simulations), that allow us to capture the asymmetries of the pulses while still maintaining purely dipolar magnetic fields. 

\begin{table}
\caption{ Best-fits parameters for the X-rays signals of the four target millisecond pulsars, when filtering-out the additional ERs over the close-region. 
}
\vspace{0.2cm}
\centering {
\begin{tabular}{ c | c  c  c    c   c  | c}         \hline \hline
~ PSR ~    & ~ $\mathcal{C}$ ~ & ~ $\chi$ ~ & ~ $\xi$ ~  & ~ $b$ ~ & ~ $\kappa \bar{\gamma}$ ~ & $\chi_{r}^2$\\ \hline
J0437–4715 &  $~ 0.22 ~$  & $~ 60 \degree ~$  & $~ 26 \degree ~$ & $~ 0.52 ~$  & $~ 7.0 \times 10^6 ~$ & $~ ~ 8.7~$ \\ 
J1231−1411 &  $~ 0.25 ~$  & $~ 15 \degree ~$  & $~ 69 \degree ~$ & $~ 0.64 ~$  & $~ 1.3 \times 10^7 ~$ & $~ ~ 3.6~$ \\
           &  $~ 0.25 ~$  & $~ 15 \degree ~$  & $~ 76 \degree ~$ & $~ 0.78 ~$  & $~ 1.3 \times 10^7 ~$ & $~ ~ 5.6~$ \\
J2124−3358 &  $~ 0.25 ~$  & $~ 15 \degree ~$  & $~ 78 \degree ~$ & $~ 0.58 ~$  & $~ 3.5 \times 10^7 ~$ & $~ ~ 3.7~$ \\           
J0030+0451 &  $~ 0.25 ~$  & $~ 45 \degree ~$  & $~ 85 \degree ~$ & $~ 0.78 ~$  & $~ 3.2 \times 10^7 ~$ & $~ 24.4~$ \\ \hline  \hline
\end{tabular}
}
\label{table:filt-fits}
\end{table}

Excluding the closed ER --as well as bringing its averaged temperature closer to the one of the open ER (like the case shown in Fig.\ref{fig:PSR0451-alt})-- implies a much narrower distribution around a single effective temperature. We notice that, within the emission model employed, this would typically lead to poorer fits of the spectral shapes.
On the other hand, we find that the spectrum of the best-fit configuration for PSR J1231−1411 excluding the closed ER (from Table \ref{table:filt-fits}) remains well described by the model.

\subsection{Normalization issues} \label{sec:normalization}

Arguably, the main limitation of our current modelling is reflected on the overall flux intensities, which --as indicated in figures \ref{fig:PSR4715-best} to \ref{fig:PSR0451-alt}-- are about $2$--$4$ orders of magnitude higher as compared to the (fitted) observational data. 

We shall examine here in some detail the case of PSR J0030+0451, for which there is more data analysis available. To this end, we take its best-fit configuration from Table~\ref{table:best-fits} (the first row for this pulsar) and estimate the effective emitting areas and the averaged effective temperatures for the two kinds of emitting regions; namely, the open/closed ERs. The resulting values are summarized on Table~\ref{table:estimations}, together with analogous parameters of a few relevant spectral fits from \cite{bogdanov2009deep} for blackbody (BB) and Hydrogen (Hatm) atmospheric models, with and without additional power-law (PL) components.

\begin{table}
\caption{ Estimated effective averaged temperatures and emitting areas in our model, compared with analogous parameters from different spectral fittings for PSR J0030+0451.\\
{\footnotesize (*) i.e.: BB with beaming. $T_{\rm eff, 1}$ and $T_{\rm eff, 2}$ represent in this case, the averaged effective temperatures over the open and closed ERs, respectively. Whereas $R_{\rm eff, 1/2}$ are estimated from the corresponding net emitting areas of each region, as $A_{\rm eff, 1/2} = 4 \pi R_{\rm eff, 1/2}^2$. }
}
\vspace{0.2cm}
\centering {
\begin{tabular}{ l | c  c  c  c }         \hline \hline
~ Spectral model ~  & ~ $T_{\rm eff, 1}$ ~ & ~ $R_{\rm eff, 1}$~ & ~ $T_{\rm eff, 2}$ ~  & ~ $R_{\rm eff, 2}$ ~ \\ 
~  ~& ~[$10^6$ K]   ~ & ~[km] ~ & ~ [$10^6$ K] ~  & ~ [km] ~ \\ \hline
This work (*) &  $~ 3.2 ~$  & $~1.17~$  & $~ 1.5 ~$ & $~ 3.55 ~$  \\  
BB(x2)   &  $~ 3.1 ~$  & $~0.05~$  & $~ 1.5 ~$ & $~ 0.28 ~$  \\ 
Hatm(x2) &  $~ 1.7 ~$  & $~0.25~$  & $~ 0.7 ~$ & $~ 2.2 ~$  \\ 
BB(x2) + PL   &  $~ 2.5 ~$  & $~0.08~$  & $~ 1.2 ~$ & $~ 0.34 ~$  \\ 
Hatm(x2) + PL &  $~ 1.4 ~$  & $~0.37~$  & $~ 0.7 ~$ & $~ 1.9 ~$  \\ \hline  \hline
\end{tabular}
}
\label{table:estimations}
\end{table}

We observe that our averaged temperatures are similar to those from the BB fits, but our emitting areas are a factor of $\gtrsim 100$ larger than the corresponding BB ones, thus explaining the roughly two orders of magnitude excess in our fluxes.
However, we notice that --since the spectrum of a Hydrogen atmosphere is harder than that of a BB for the same effective temperature-- the emitting areas inferred for PSR J0030+0451 with the Hatm model are much closer to ours (although still smaller,  by a factor $\sim 3$). On the other hand, we note that the effective temperatures are lower in the Hatm model, but still maintaining the same proportion, $T_{\rm eff, 1} \approx 2 T_{\rm eff, 2}$. 

Therefore, we might expect that replacing the BB (with beaming) in our emission model by an Hydrogen atmosphere, and by re-adjusting the $\kappa \bar{\gamma}$ values accordingly (i.e., reducing the effective temperatures by a factor $\sim 1/2$), we should get much more consistent spectra in absolute physical units for all our target pulsars, while keeping the good quality of their light curves fits. Testing this hypothesis is, however, beyond the scope of the present article and we defer it to a future work.

Following the above reasoning, we could look back to the spectrum found in Fig.~\ref{fig:PSR0451-alt} for the alternative best-fit of PSR J0030+0451. We note that by increasing the value of $(\kappa \bar{\gamma})_{\rm closed}$ on a factor $\sim 4$, one is increasing the effective temperature of this ER by a factor $\sim 1.4$; thus, bringing it closer to the averaged values of the hotter (open) region.
This has shown a negative impact for fitting the spectrum with our BB $+$ beaming model, but it would not necessarily be the case with the Hatm model. For instance, one of the most favoured synthetic ER models constructed in \cite{riley2019}, ST+PST, consist on a single effective temperature along with the Hatm atmospheric model. 

A more thorough exploration of such decoupling for $\kappa \bar{\gamma}$ in the emission model, both from a physical perspective and from its implications for fitting the light curves, as well as shifting to an Hydrogen atmospheric model, will be a clear continuation route for a future study.


\section{Conclusions}\label{sec:conclusions}

In this article, we modelled thermal X-ray signals produced by MSPs and fitted them to observational data from several target pulsars: PSR J0437–4715, PSR J1231−1411, PSR J2124−3358 and PSR J0030+0451. 
Our approach starts by numerically solving the pulsar magnetospheres under the FF approximation,
assuming a simple centered dipolar magnetic field. The solutions are then linked to an effective emissivity over the stellar surface trough a simple emission model; followed by the propagation of the emitted photons (via ray tracing) to compute light curves and spectra. These are finally compared against the observational data of the target MSPs. 

We found non-standard emission regions (ERs) on the stellar surface, associated to spacelike electric currents over the closed-zone of the pulsar. Such ERs appear as a consequence of having included the gravitational curvature of the NS in the FF simulations, and they are also related to the null charge surfaces in the magnetosphere.  

With these closed ERs included in our emission model, we were able to achieve excellent fits to the light curves for all the four MSPs considered. Simultaneously, we have obtained a good agreement with the spectral shape of their thermal emission. This is quite remarkable, given that these X-rays light curves have been typically attributed to the presence of strong higher-multipole moments and/or off-centered dipolar components in the magnetic field (e.g., \cite{bogdanov2007, bogdanov2008, bogdanov2012, lockhart2019x}), especially for the widely studied case of PSR J0030+0451 \cite{bilous2019, chen2020,kalapotharakos2021}. 

We have observed that without these additional (closed) ERs, the light curves fitting become significantly worse, even at the qualitative level. The reason for this, is that the closed ERs provides an additional contribution to the signal, capable of producing the asymmetries found in between the main pulses (interpulses) of the observed light curves. That is, even when the individual signals arising from the (antipodal) open/closed ERs are quite symmetric, the slight dephasing among them lead to the necessary interpulses modulations when combined.  

The spirit of this work is not to claim we have found the ``correct" model, but rather to emphasize the need for a better connection between the global magnetospheric properties and the determination of the emitting regions, as well as the appropriate modelling of such emissions.
We have shown that including the spacetime curvature on the magnetosphere's description results in more complex ERs, that might account for the observed light curves without the need of invoking large non-dipolar magnetic field components and/or offsets from the NS center. In particular, we showed that (regardless of the particular emission model employed here) antipodal ERs can indeed produce X-ray light curves approaching those detected by NICER. This also raises interesting questions regarding the physical nature of these additional spacelike currents and their entailment to the bombardment mechanism responsible for heating the stellar surface. Will these currents contribute in the same way to the bombardment as those reaching the polar caps along open field lines? Should their thermal emissions be modelled separately?

We have recognized that the emission model considered is the main limitation of our present X-ray modelling, which considerably overestimates the observational fluxes.  For the BB+beaming model assumed here, the effective emitting areas of our ERs turn-out to be too large. However, we noticed that for an Hydrogen atmospheric model --which has a much harder spectrum as compared to BB-- the situation would improve significantly. We expect that an Hatm model should lead to more consistent spectra in absolute physical units, while keeping the good quality of the light curve fits found here for all the target MSPs. 

A natural next step to pursue this research further, will be to improve the emission model by considering an Hatm model with a more realistic (self-consistent) angular distribution (i.e., beaming pattern), as well as a more refined modelling of the bombardment mechanism (based on the FF currents) along the lines of \cite{baubock2019, salmi2020}. Such refinement, could incorporate the idea of decoupling the description of the two type of ERs, namely: open/closed ERs.
Moreover, a more sophisticated simultaneous fitting based on the full phase-resolved spectral data of PSR J0030+0451 would be an interesting improvement of this study. As well as expanding the application of our model to other MSPs, if sufficiently accurate X-ray data is available.


\section{Acknowledgments}

The authors would like to thank Rosalba Perna for helpful suggestions. F.C., J.P. and O.R acknowledge financial support from CONICET, SeCyT-UNC, and MinCyT-Argentina. D.V. is funded by the European Research Council
(ERC) under the European Union’s Horizon 2020 research
and innovation programme (ERC Starting Grant IMAGINE,
No. 948582) and his institution is supported by the Spanish
program Unidad de Excelencia Mar{\i}a de Maeztu, CEX2020-
001058-M. C.P. is supported by the Grant PID2019-110301GB-I00 funded by MCIN/AEI/10.13039/501100011033 and by "ERDF A way of making Europe".
Numerical computations were performed on the Sakura cluster at Max-Planck Computing and Data Facility, and on the Serafin Cluster at Centro de Computación de Alto Desempeño, Universidad Nacional de Cordoba, which is part of the Sistema Nacional de Computación de Alto Desempeño, MinCyT-Argentina.


\appendix

\section{Light curves best-fits search and posterior probability density distributions} \label{sec:appendix}

\begin{figure*}
    \subfloat[ PSR J0437-4715 \label{cones:a}]{%
      \includegraphics[scale=0.56]{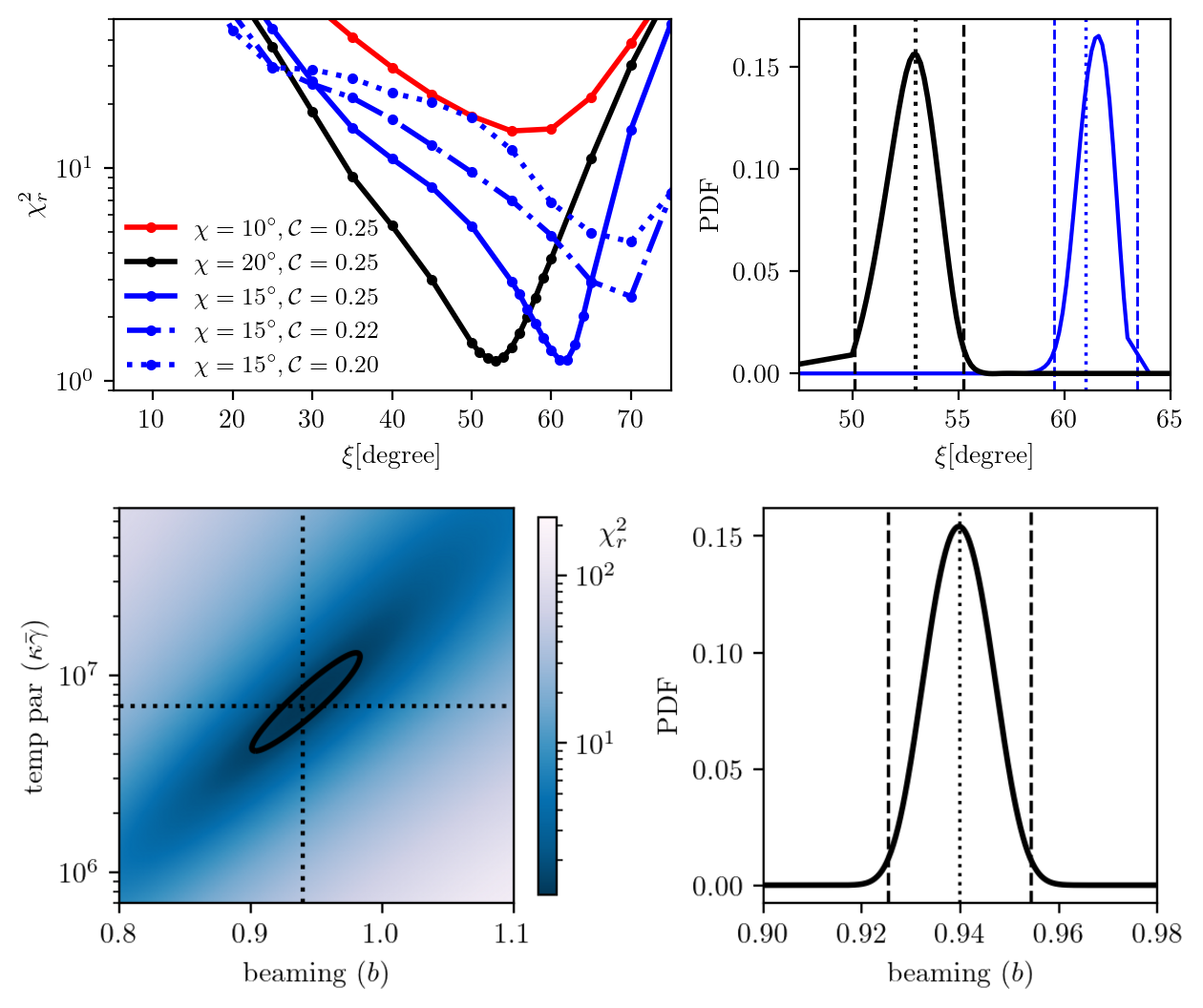}
    }
    \subfloat[PSR J1231-1411 \label{cones:b}]{%
      \includegraphics[scale=0.56]{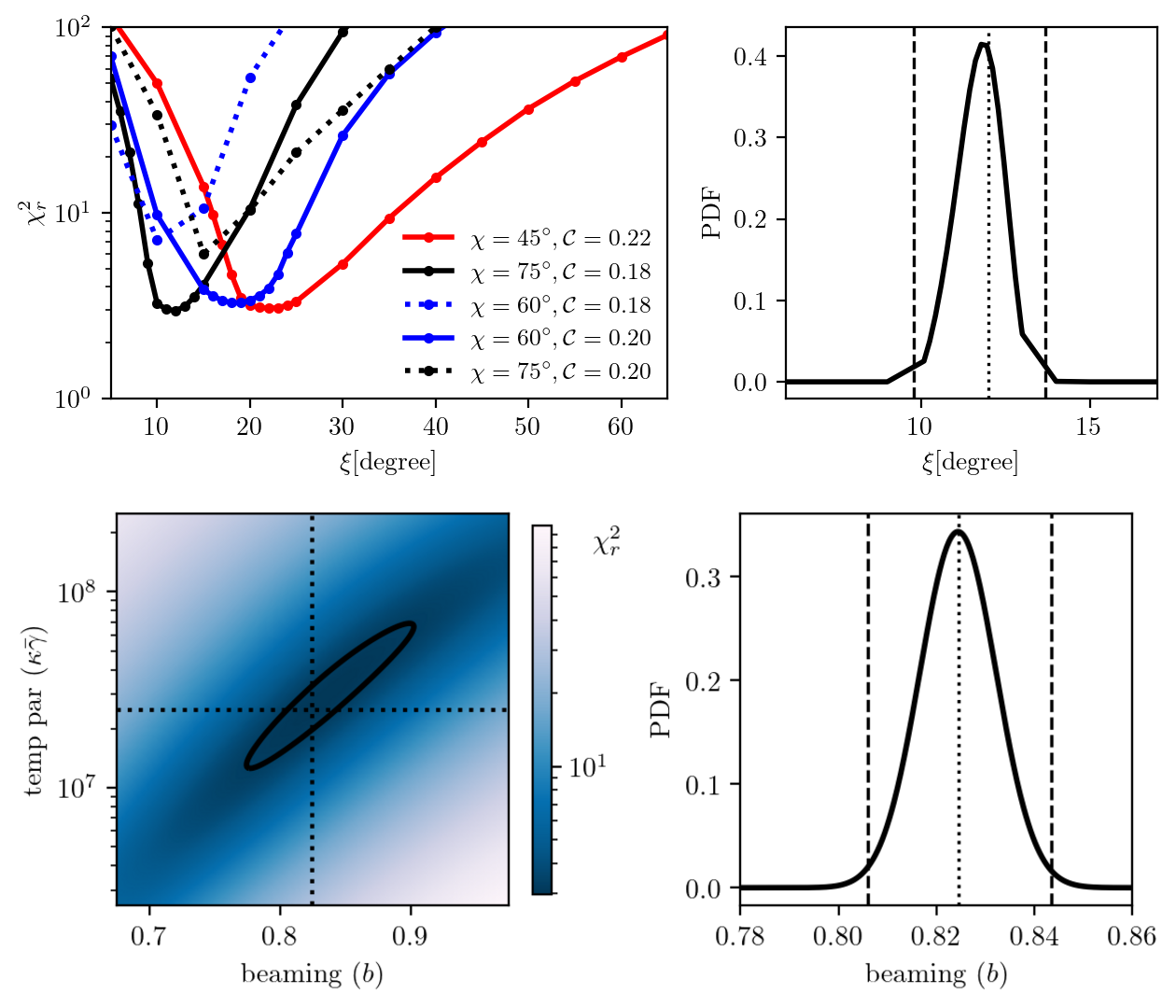}
    }
    \vspace{0.5cm}
    \subfloat[PSR J2124-3358 \label{cones:c}]{%
      \includegraphics[scale=0.56]{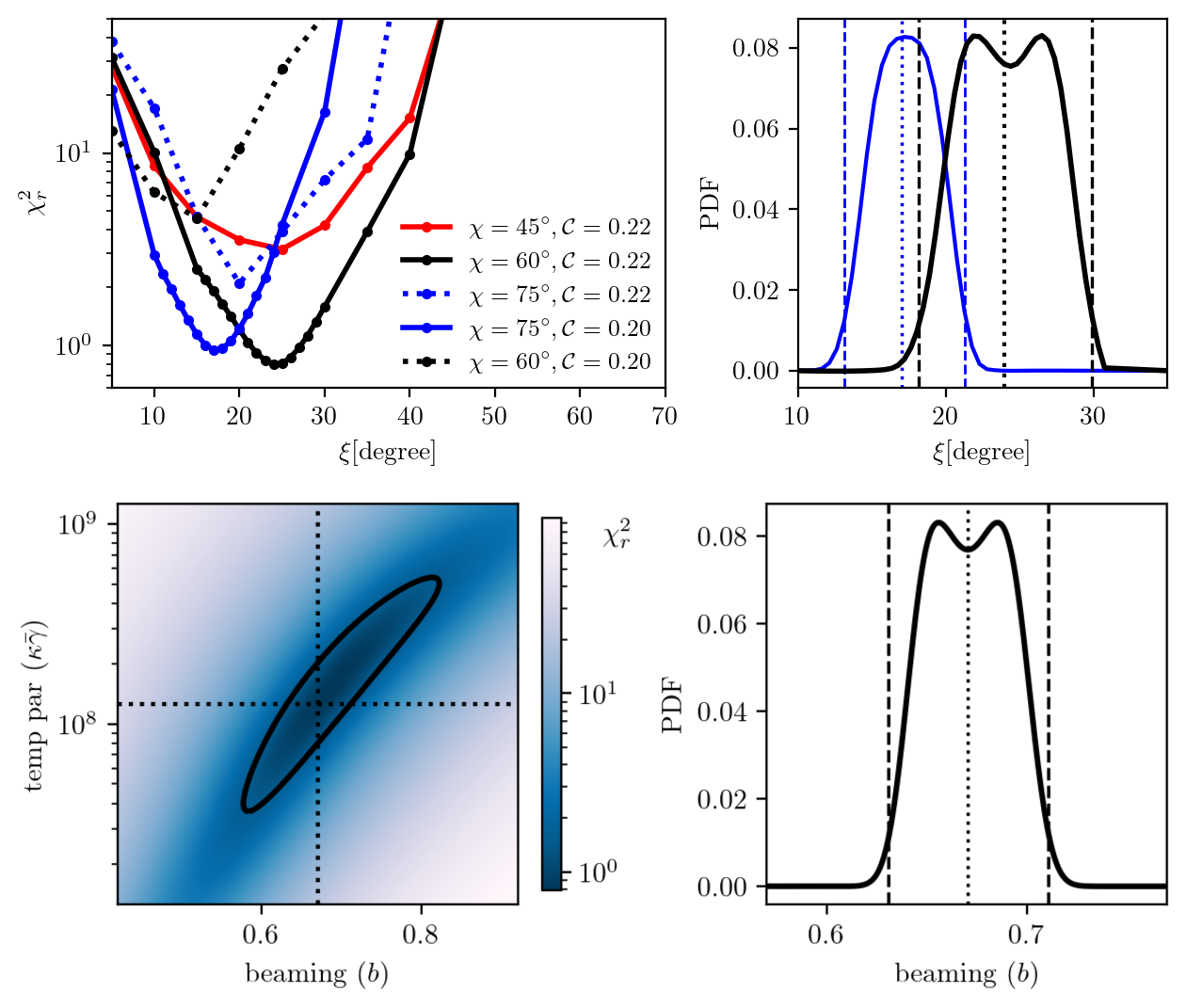}
    }
    \subfloat[PSR J0030+0451 \label{cones:d}]{%
      \includegraphics[scale=0.56]{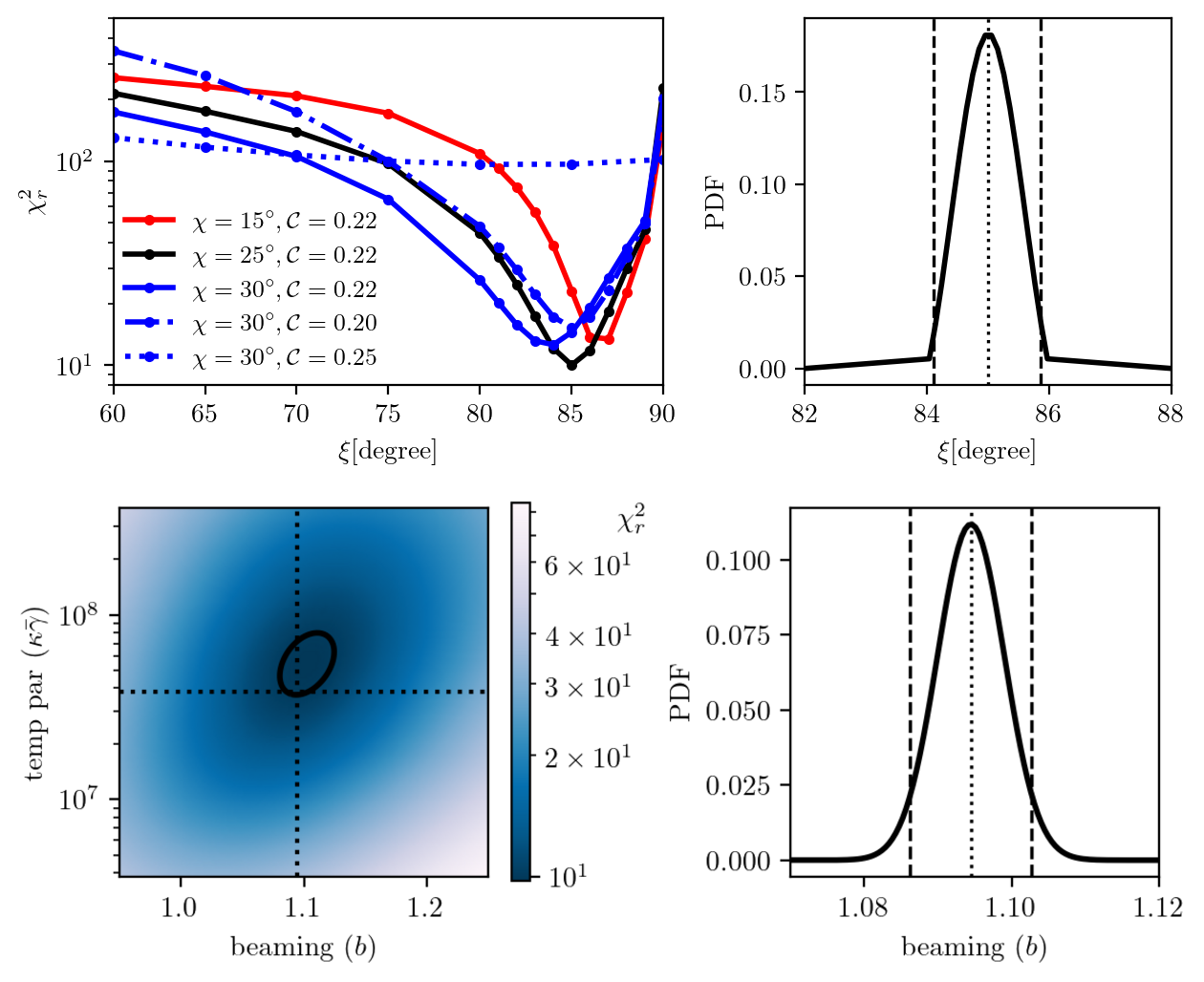}
    }
    \vspace{1cm}    
    \caption{\textit{Light curves best-fit search and posterior probability distributions for each of the target pulsars (a), (b), (c) and (d).}
    \textbf{Top-left plots:} $\chi_{r}^2$ against $\xi$, displayed for various representative pairs ($\mathcal{C}$, $\chi$) around the best-fit configurations. The curves in solid-black depict the "main" best-fits (i.e., first row of each MSP on Table~\ref{table:best-fits}). \textbf{Top-right plots:} Posterior PDFs on the viewing angle $\xi$. Vertical dotted lines indicate central (discrete) values of $\xi$ minimizing $\chi_{r}^2$, while dashed ones display the $95$\% credible intervals. Colors follows consistently from the top-left plots.
    \textbf{Bottom-left plots:} 2D posterior distribution in $\kappa \bar{\gamma}$ and $b$ (fixing all the remaining model parameters to their "main" best-fit values). The colorscale represents $\chi_{r}^2$; dotted vertical and horizontal lines mark the central values of $\kappa \bar{\gamma}$ and $b$; while the solid-black curve enclose the $95$\% credible regions. \textbf{Bottom-right plots:} Posterior PDFs on the beaming parameter $b$.
    }
    \label{fig:search}
  \end{figure*}

In this section, we provide with some further details on the search for the light curves best-fits presented on Sec.~\ref{sec:LCandS} and, in particular, we analyse the posterior distributions for the most relevant parameters in our model.
This information, which is summarized in Fig.~\ref{fig:search}, might help to illustrate the $\chi_{r}^2$ minimization procedure, while also serving as indication for the quality of the fits.

From the $7$ parameters of our complete model, $X=\{ \mathcal{C}, \chi, \xi, b, \kappa \bar{\gamma}, \varphi_o , \lambda_o \}$: one ($\kappa \bar{\gamma}$) is fixed at adjusting the spectral distributions and is not actually being varied when fitting the light curves; two of them ($\varphi_o , \lambda_o$) are not physically relevant in the present context\footnote{ Although they could acquire more significance on a more detailed or complementary analysis, as for instance if the absolute azimuth of the dipole moment can be co-related with the presumed zero-phase point (e.g., from radio or gamma-rays signals) of the X-ray light curve.}; while $\mathcal{C}$ and $\chi$ are discrete and limited by the computational cost of the GRFF simulations. This is the reason why we have focused here mainly on the posterior probability distributions of the beaming index $b$ and viewing angle $\xi$ (also discrete, but only limited by the much lower cost of the ray tracing).

For given observational data $\mathcal{O}$ and model $X$, we adopt a Bayesian approach (see, e.g., \cite{bernardo2009}) with no assumed priors, for which the posterior probability distribution is given by,
\begin{equation}
    P(X|\mathcal{O}) = \frac{P(\mathcal{O}|X)}{P(\mathcal{O})} 
\end{equation}
where $P(\mathcal{O}|X)$ represents the \textit{likelihood function} and $P(\mathcal{O})$ is called \textit{evidence}. For each (available) point in our parameter space, we can compute its associated \textit{chi-square} value from equation \eqref{eq:dist}. We then assume a \textit{chi-square} distribution to get the \textit{likelihood function},
\begin{equation}
    P(\mathcal{O}|X) = f(\chi^2, \varpi) := \frac{(\chi^2)^{\varpi/2 -1} \, e^{-\chi^{2}/2}}{2^{\varpi/2} \, \Gamma(\varpi/2)} 
\end{equation}
where $\Gamma(\nu/2)$ is the \textit{Gamma function} and we recall that $\varpi=n-k$ (being $n$ the number of phase-bins on the data $\mathcal{O}$ and $k=\#X$ the number of parameters in the model).

Hence, we shall consider the posterior probability distribution function (PDF) to be the above \textit{likelihood function} normalized by \textit{evidence},
\begin{equation}
    P(\mathcal{O}) := \int P(\mathcal{O}|X) \, dX = \int_{\chi_{\rm min}^2}^{\infty} f(\chi^2, \varpi) \, d\chi^2
\end{equation}
where $\chi_{\rm min}^2$ is the minimum \textit{chi-square} value attained for a given MSP.  
The $95$\% credible intervals shown in the plots are, thus, obtained from $\chi_{\alpha}^2$ given by the expression,
\begin{equation}
    \int_{\chi_{\alpha}^2}^{\infty} f(\chi^2, \varpi) \, d\chi^2 = \alpha \int_{\chi_{\rm min}^2}^{\infty} f(\chi^2, \varpi) \, d\chi^2
\end{equation}
for a value of $\alpha = 0.05$.

Figure \ref{fig:search} displays the \textit{reduced chi-square} as a function of $\xi$ (top-left plots) for a few relevant pairs ($\mathcal{C}$, $\chi$) and as a function of $b$ (bottom-left plots), around the minimal $\chi_{r}^2$ configurations for each target MSP. Whereas the right plots present the posterior PDF for $\xi$ and $b$, respectively\footnote{ Notice that these are not the marginalized posteriors (which involve integration of the PDF over the remaining parameters), but rather the posterior PDF evaluated by fixing all the other model parameters to their best-fits values.}.  

For PSR J0437-4715 and PSR J2124-3358, their two best-fit configurations from Table~\ref{table:best-fits} are simultaneously represented on the posterior distribution of $\xi$, further illustrating the degeneracy's arising on the geometric parameters $\mathcal{C}$, $\chi$ and $\xi$. This fact is also reflected on the top-left plot of PSR J1231-1411 (although its minima is very likely not a global one, as discussed on Sec.~\ref{sec:LCandS}).

We observe that the posterior PDF in the case of PSR J2124-3358 does not exhibit a maximum exactly at their best-fit configurations, which is expected since they have $\chi_{r}^2 < 1$ and the assumed $\chi^2$--distribution peaks at $\chi_{r}^2 = 1$. 

In the case of PSR J0030+0451, we see that the value of $\kappa \bar{\gamma}$ chosen to adjust the spectral distribution is not the most favorable for fitting the light curve. Even at the fixed pair $(\lambda_o , \varphi_o )$ associated to this $\kappa \bar{\gamma}$, there are other configurations with lower values of $\chi_{r}^2$ in the 2D posterior distribution of Fig.~\ref{fig:search}(d). Thus, indicating that a larger contribution from the closed ER would be favored, which is consistent with the alternative fit showed in Fig.~\ref{fig:PSR0451-alt}.


\bibliographystyle{unsrt} 
\bibliography{FFE}


\end{document}